\documentclass[aps,prl,reprint,superscriptaddress]{revtex4-1}

\usepackage{graphicx}
\usepackage{amssymb, amsmath}
\usepackage[usenames]{color}
\usepackage{dcolumn}
\usepackage{bm}
\usepackage{ textcomp }

\begin{document}
\title{Quantum Spin Hall Effect and Spin Bott Index in Quasicrystal Lattice}

\author{Huaqing Huang}
\affiliation{Department of Materials Science and Engineering, University of Utah, Salt Lake City, Utah 84112, USA}

\author{Feng Liu\footnote{Corresponding author: fliu@eng.utah.edu}}
\affiliation{Department of Materials Science and Engineering, University of Utah, Salt Lake City, Utah 84112, USA}
\affiliation{Collaborative Innovation Center of Quantum Matter, Beijing 100084, China}

\date{\today}

\begin{abstract}
Despite the rapid progresses in the field of quantum spin Hall (QSH) effect, most of the QSH systems studied up to now are based on crystalline materials. Here we propose that the QSH effect can be realized in quasicrystal lattices (QLs). We show that the electronic topology of aperiodic and amorphous insulators can be characterized by a spin Bott index $B_s$. The nontrivial QSH state in a QL is identified by a nonzero spin Bott index $B_s=1$, associated with robust edge states and quantized conductance. We also map out a topological phase diagram in which the QSH state lies in between a normal insulator and a weak metal phase due to the unique wavefunctions of QLs. Our findings not only provide a better understanding of electronic properties of quasicrystals but also extend the search of QSH phase to aperiodic and amorphous materials that are experimentally feasible.
\end{abstract}


\maketitle

\paragraph{Introduction.}
QSH states, which are characterized by topologically protected metallic edge states with helical spin polarization residing inside an insulating bulk gap, have attracted much interest in the last decade \cite{PhysRevLett.95.226801, PhysRevLett.95.146802, PhysRevLett.96.106802, BHZ,konig}. QSH states have been discovered in various two-dimensional (2D) materials \cite{huanghqCMS, wang2017computational}. Recently, QSH systems grown on conventional semiconductor substrates have been theoretically proposed and experimentally realized, \cite{zhou2014epitaxial,zhou2014formation,hsu2015nontrivial, wang2016quantum, reis2017bismuthene} which bridges topological states with conventional semiconductor platforms offering an attractive route towards future quantum device applications. However, almost all the existing QSH states are based on crystals and identified by the $\mathbb{Z}_2$ topological invariant that is defined only for periodic systems \cite{PhysRevLett.95.146802, PhysRevB.74.195312}. In this Letter, we extend the QSH state to quasicrystalline systems and define a new topological invariant, the spin Bott index, to identify nontrivial topology in aperiodic and amorphous systems.

Quasicrystal phases with long-range orientational order but no translational symmetry have attracted considerable attentions since their first discovery in 1982 \cite{PhysRevLett.53.1951}. Because of the unique structural characteristics, quasicrystals exhibit various unusual physical properties, such as extremely low friction \cite{dubois1991quasicrystalline}, self-similarity \cite{PhysRevB.34.3904,PhysRevB.37.2797} and critical (power-law decay) behaviors of wavefunctions \cite{PhysRevB.43.8879, PhysRevB.43.8892} {in between extended Bloch states of periodic systems and exponentially localized states of disorder systems due to Andersion localization.} Additionally, 2D QLs have been epitaxially grown on quasicrystalline substrates recently \cite{jenks1998quasicrystals, mcgrath2002quasicrystal, sharma2007quasicrystal, thiel2008quasicrystal, mcgrath2012memory,ledieu2014surfaces}. Here we answer an intriguing question: is it possible to realize a QSH state in a quasicrystal to support topologically protected extended edge states along the quasicrystal's boundary, contrary to the general critical states of quasicrystals? If so, how to define a topological invariant in analogue of the $\mathbb{Z}_2$ index of periodic systems, to identify the QSH state in quasicrystals and more generally in amorphous systems?

We note that so far only very few works have been done on topological states of aperiodic systems, in particular, the Chern insulator states analogous of quantum anomalous Hall (QAH) states \cite{PhysRevX.6.011016,mitchell2018amorphous,PhysRevLett.118.236402}. However, they are based on either artificial systems, such as optical waveguide with an artificial gauge field \cite{PhysRevX.6.011016} and networks of interacting gyroscopes \cite{mitchell2018amorphous}, or lattice models with artificial hoppings, such as an amorphous lattice model \cite{PhysRevLett.118.236402}. In contrast, we propose the realization of the QSH state in real quasicrystal materials with atomic orbitals, especially the surface-based QLs, which are readily makable by experiments. We characterize its nontrivial topology using the newly derived spin Bott index. As in crystals, the QSH phase in quasicrystals manifests also with robust metallic edge states and quantized conductance. Moreover, we determine a topological phase diagram for QLs and show that the QSH state lies in between a normal insulator and a gapless phase exhibiting weak metallic behavior due to the critical wavefunctions of QLs. Our proposed atomic model of quasicrystalline QSH states can be possibly grown by deposition of atoms on the surfaces of quasicrystal substrates.

\begin{figure}
\includegraphics[width =1\columnwidth]{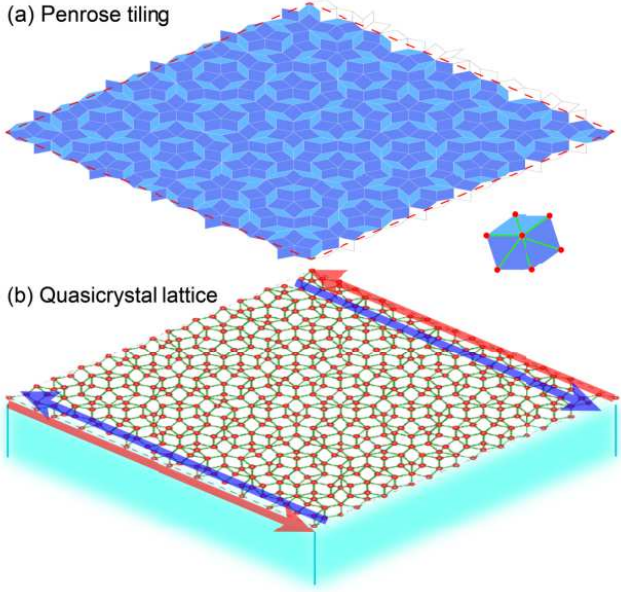}%
\caption{\label{fig1} (a) Penrose tiling containing 521 vertices. The red dashed line defines a unit cell under periodic approximation. The inset shows that the atomic orbitals placed on vertices of rhombuses and the nearest-neighbor hopping between them. (b) Atomic model of QSH state in a surface-based 2D QL. The red and blue arrows represent edge states with opposite spin polarizations.}
\end{figure}

\paragraph{Model.} 2D QLs are constructed according to the Penrose tiling with fivefold rotational symmetry \cite{penrose1974role}, as shown in Fig.~\ref{fig1}(a). Since the QL possesses long-range orientational order but lacks translational symmetry, we cannot use the Bloch theorem as for the crystal calculations. However, it is still possible to generate a series of periodic lattices with a growing number of atoms that approximate the infinite QL according to the quasicrystal tiling approximants \cite{tsunetsugu1986eigenstates,entin1988penrose, PhysRevB.43.8879,PhysRevX.6.011016}. In our model, the atoms are located on vertices of rhombuses of the Penrose tiling. The first three nearest neighbors are considered, as shown in the inset of Fig.~\ref{fig1}(a); their separations are $r_0:r_1:r_2=2\cos\frac{2\pi}{5}: 1: 2\sin\frac{\pi}{5}$, respectively.

We consider a general atomic-basis model for QLs with three orbitals ($s,p_x,p_y$) per site. The Hamiltonian is given by
\begin{eqnarray}
H&=&\sum_{i\alpha}\epsilon_\alpha c_{i\alpha}^\dag c_{i\alpha}+\sum_{<i\alpha,j\beta>}t_{i\alpha,j\beta}c_{i\alpha}^\dag c_{j\beta}\nonumber\\
 &+&i\lambda\sum_i(c_{ip_y}^\dag \sigma_z c_{ip_x}-c_{ip_x}^\dag \sigma_z c_{ip_y}),
\label{eq1}
\end{eqnarray}
where $c_{i\alpha}^\dag=(c_{i\alpha\uparrow}^\dag,c_{i\alpha\downarrow}^\dag)$ are electron creation operators on the $\alpha(=s,p_x,p_y)$ orbital at the $i$-th site. $\epsilon_\alpha$ is the on-site energy of the $\alpha$ orbital. The second term is the hopping term where $t_{i\alpha,j\beta}=t_{\alpha,\beta}(\mathbf{d}_{ij})$ is the hopping integral which depends on the orbital type ($\alpha$ and $\beta$) and the vector $\mathbf{d}_{ij}$ between sites $i$ and $j$. $\lambda$ is the spin-orbit coupling (SOC) strength and $\sigma_z$ is the Pauli matrix.
In our model, the hopping integral follows the Slater-Koster formula \cite{SlaterKoster}
\begin{equation}
t_{\alpha,\beta}(\mathbf{d}_{ij})=\mathrm{SK}[V_{\alpha\beta}(d_{ij}),\hat{\mathbf{d}}_{ij}],
\end{equation}
where $\hat{\mathbf{d}}_{ij}$ is the unit direction vector. The distance dependence of the bonding parameters $V_{\alpha\beta}(d_{ij})$ is captured approximately by the Harrison relation \cite{harrison2012electronic}:
\begin{equation}
V_{\alpha\beta}(d_{ij})=V_{\alpha\beta,0}\frac{d_0^2}{d_{ij}^2},
\end{equation}
where $d_0$ is a scaling factor to uniformly tune the bonding strengths. Since only the band inversion between \textit{s} and \textit{p} states of different parities is important for the realization of topological states, we focus mainly on 2/3 filling of electron states hereafter, unless otherwise specified.

\begin{figure}
\includegraphics[width =1\columnwidth]{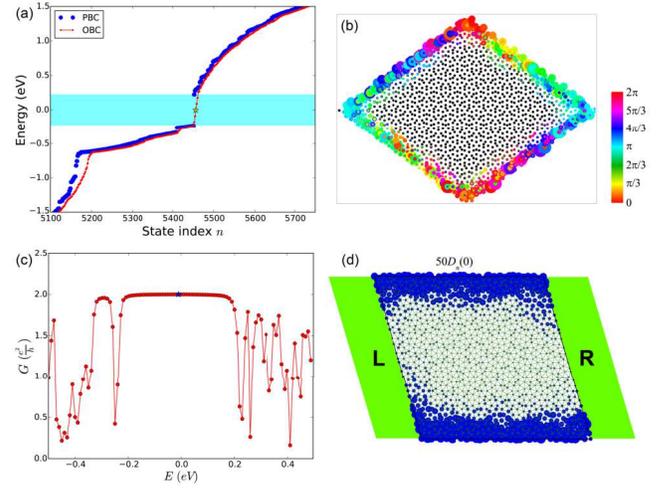}%
\caption{\label{fig2} Calculation of a quasicrystal sample with 1364 atoms (3 orbitals and 2 spins on each atom). The parameters used here are $\epsilon_s=1.8, \epsilon_{p}=-6.5, \lambda=0.8, V_{ss\sigma}=-0.4, V_{sp\sigma}=0.9,V_{pp\sigma}=1.8$ and $V_{pp\pi}=0.05$ eV. (a) Energy eigenvalues $E_n$ versus the state index $n$. The system with periodic boundary condition (PBC) shows a gap, while that with open boundary condition (OBC) shows mid-gap states. (b) The wavefunction $|\psi(\mathbf{r})\rangle=\rho(\mathbf{r})e^{i\phi(\mathbf{r})}$ of the mid-gap state [marked as the yellow star in (a)] is localized on the edge of the system. The size and the color of the blob indicates the norm $|\rho(\mathbf{r})|^2$ and phase $\phi(\mathbf{r})$ of the wavefunction, respectively. (c) Two-terminal conductance $G$ as a function of the Fermi energy $E$ showing a quantized plateau in the energy gap. (d) Local density of state $D_n(E)$ at $E=0$ eV for the central quasicrystal in the transport simulation. The size of blue dot represents the relative value of local density of state.}
\end{figure}

\paragraph{Results.} The calculated results of a particular realization of the QSH state in the QL with open boundary condition (OBC) and periodic BC (PBC) are shown in Fig.~\ref{fig2}. It is found that the PBC system clearly shows an energy gap, as displayed in Fig.~\ref{fig2}(a). This indicates that the system is an insulator. However, there are some eigenvalues within the gap region of the energy spectrum in the presence of OBC, implying that the system becomes metallic. In Fig.~\ref{fig2}(b), we plotted the wavefunction distribution of a typical mid-gap state [marked as a star in Fig.~\ref{fig2}(a)]. Interestingly, the mid-gap state is an ``edge state'' which is localized on the boundary of the finite QL. We also studied other finite QL samples with different boundary geometries, and found that the edge state is robust, which always remains on the boundaries regardless of their detailed shapes \footnote{\label{fn} See the joint publication: H. Huang and F. Liu, Phys. Rev. B xx,xxx (2018), for more details.}. In contrast, typical bulk states show the well-known localized or critical characters of quasicrystals \footnotemark[\value{footnote}]. Due to the time-reversal symmetry, the eigenvalues always appear in pairs with the same energy for the mid-gap edge states; while the wavefunctions of the two ``degenerate'' states are mainly contributed from spin up and down components, respectively. This implies that the system should be a topologically nontrivial QSH insulator.

To verify the conductive feature of the edge states, we further investigated the transport properties based on the non-equilibrium Green's function method \cite{datta1997electronic,PhysRevB.38.9375,huanghqInterface}. In the transport simulation, one finite QL is coupled to two semi-infinite periodic leads and the two-terminal conductance is calculated as shown in Fig.~\ref{fig2}(c). Remarkably, there is a clear quantized plateau at $G =2e^2/h$ for the two-terminal charge conductance, which resembles that of the QSH state in graphene \cite{PhysRevLett.95.226801}. As shown in Fig.~\ref{fig2}(d), the local density of state of the central quasicrystal at $E=0$ eV [the blue star marked in Fig.~\ref{fig2}(c)] mainly distributes on two edges of the quasicrystal, indicating that the conductive channels are mostly contributed by the topological edge states.

The bulk energy gap, robust mid-gap edge states as well as the quantized conductances connote the nontrivial topology of the QL. However, the most critical quantity to identify the electronic topology is the topological invariant which classifies insulators into different topological classes. For example, the Chern number ($C$) \cite{Haldane,chang2013experimental,huanghqSemiDirac} distinguishes the QAH states ($C\neq 0$) with trivial time-reversal-broken insulators ($C=0$); the topological $\mathbb{Z}_2$ invariant \cite{PhysRevLett.95.146802,PhysRevB.74.195312} determines the QSH states ($\mathbb{Z}_2=1$) with normal time-reversal-invariant insulators ($\mathbb{Z}_2=0$). However, these topological invariants are only applicable to periodic systems. Recently, the Bott index, which is equivalent to the Chern number \cite{toniolo2017equivalence}, is proposed to determine QAH state in nonperiodic system \cite{loring2011disordered, hastings2011topological,loring2015k, PhysRevLett.118.236402, PhysRevX.6.011016}. Here we derive a topological invariant for the QSH state of aperiodic systems.

\paragraph{Spin Bott index.} In order to verify the QSH states in the QLs, we define the spin Bott index for QSH states (For a detailed discussion see \footnotemark[\value{footnote}].) in reference to the definition of the spin Chern number \cite{PhysRevLett.97.036808,PhysRevB.75.121403, PhysRevB.80.125327,*prodan2010non,*prodan2011disordered}. We firstly construct the projector operator of the occupied states,
\begin{equation}
P=\sum_i^{N_{occ}}|\psi_i\rangle\langle\psi_i|.
\end{equation}
Then we make a smooth decomposition $P_z=P_+\oplus P_-$ for spin-up and spin-down sectors. At the first sight, this seems an easy job as long as two spin channels are decoupled. However, it becomes more complicated if there are spin-mixing  terms in the Hamiltonian. The key idea is to use the projected spin operator,
\begin{equation}
P_z=P\hat{s}_zP,
\end{equation}
where $\hat{s}_z=\frac{\hbar}{2}\sigma_z$ is the spin operator ($\sigma_z$ is the Pauli matrix). For a spin conserving model, $\hat{s}_z$ commutes with the Hamiltonian $H$ and $P_z$, the Hamiltonian as well as eigenvectors can be divided into spin-up and spin-down sectors. Therefore, the eigenvalues of $P_z$ consist of just two nonzero values $\pm\frac{\hbar}{2}$. For systems without spin conservation (for example, the Kane-Mele model with a nonzero Rashba term $\lambda_R$ \footnotemark[\value{footnote}]), the $\hat{s}_z$ and $H$ no longer commute, and the spectrum of $P_z$ spreads toward zero. However, as long as the spin mixing term is not too strong (e.g., $\lambda_R/\lambda_{SO}<3$ in the Kane-Mele model, where $\lambda_{SO}$ is the intrinsic spin-orbit coupling \footnotemark[\value{footnote}]), the eigenvalues of $P_z$ remain two isolated groups which are separated by zero. Because the rank of matrix $P_z$ is $N_{occ}$, the number of positive eigenvalues equals to the number of negative eigenvalues, which is $N_{occ}/2$. The corresponding eigenvalue problem can be denoted as
\begin{equation}
P_z|\pm \phi_i\rangle=S_\pm|\pm\phi_i\rangle.
\end{equation}
In this way we can construct new projector operators
\begin{equation}
P_\pm=\sum_i^{N_{occ}/2}|\pm\phi_i\rangle\langle\pm\phi_i|,
\end{equation}
for two spin sectors. Next, we calculate the projected position operators
\begin{eqnarray}
U_\pm=P_\pm e^{i2\pi X}P_\pm+(I-P_\pm),\\
V_\pm=P_\pm e^{i2\pi Y}P_\pm+(I-P_\pm),
\end{eqnarray}
where $X$ and $Y$ are the rescaled coordinates which are defined in the interval $[0,1)$. To make the numerical algorithm more stable, we perform a singular value decomposition (SVD) $M=Z\Sigma W^\dag$ for $U_\pm$ and $V_\pm$, where $Z$ and $W$ are unitary and $\Sigma$ is real and diagonal, and set $\tilde{M}=ZW^\dag$ as the new unitary operators. The SVD process does not destroy the original formalism, but effectively improves the convergence and stability of the numerical algorithm \footnotemark[\value{footnote}]. The Bott index, which measures the commutativity of the projected position operators \cite{bellissard1994noncommutative, hastings2010almost, exel1991invariants, katsura2016Z2,*katsura2018noncommutative}, are given by
\begin{equation}
B_\pm=\frac{1}{2\pi}\textrm{Im}\{\textrm{tr}[\log(\tilde{V}_\pm \tilde{U}_\pm \tilde{V}_\pm^\dag \tilde{U}_\pm^\dag)]\},
\end{equation}
for two spin sectors, respectively.
Finally, we define the spin Bott index as the half difference between the Bott indices for the two spin sectors
\begin{equation}
B_s=\frac{1}{2}(B_+-B_-).
\end{equation}
We checked the above definition for crystalline and disorder systems and found that the spin Bott index is the same as the $\mathbb{Z}_2$ topological invariant \footnotemark[\value{footnote}].

Similar to the spin Chern number \cite{PhysRevLett.97.036808,PhysRevB.75.121403,PhysRevB.80.125327,*prodan2010non,*prodan2011disordered}, the spin Bott index is a well-defined topological invariant. It is applicable to quasiperiodic and amorphous systems, which provides a useful tool to determine the electronic topology of those systems without translational symmetry. For the QL in Fig.~\ref{fig2}, we found that the spin Bott index $B_s = 1$, indicating indeed a QSH state. Due to the bulk-edge correspondence, it is natural to expect the existence of robust boundary states for systems with nontrivial spin Bott index. Hence, the nontrivial spin Bott index is consistent with the above calculations of edge states and electronic conductance, all confirming the nontrivial topological character of the QL.

\begin{figure}
\includegraphics[width =1\columnwidth]{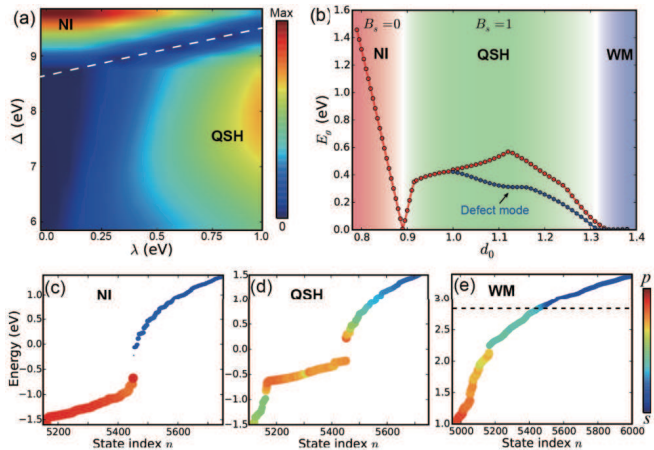}%
\caption{\label{fig3} (a) Topological phase diagram the in the parameter space of energy difference $\Delta=\epsilon_s-\epsilon_p$ and SOC strength $\lambda$. (b) Energy gap $E_g$ and spin Bott index $B_s$ as a function of interaction strength scale $d_0$ calculated in a quasicrystal approximant containing 3571 atoms. A topological phase transition among normal insulator (NI), QSH insulator and weak metal (WM) is clearly visible. The dark blue line in the QSH region represents the defect mode in the energy gap. (c,d,e) Orbital-resolved spectrum of QL in (c) NI, (d) QSH insulator and (e) WM states, respectively. A \textit{s}-\textit{p} band inversion occurs in the QSH state. The color of dots represents the relative weight of $p$ or $s$ orbitals.
}
\end{figure}

\paragraph{Topological phase diagram.} In order to achieve the QSH state, one prerequisite is the band inversion between conduction and valance states. Generally, one can realize the band inversion by tuning either the on-site energy difference $\Delta=\epsilon_s-\epsilon_p$ or the SOC strength $\lambda$ \cite{huanghqCMS,wang2017computational,wang2016quantum}. To investigate the necessary condition for the realization of topological states in QLs, we calculated the topological phase diagram in the $\Delta$-$\lambda$ plane. As shown in Fig.~\ref{fig3}(a), the normal insulator (NI) and QSH state are divided by an energy gap closing line [white dashed line in Fig.~\ref{fig3}(a)]. To achieve the QSH phase, one has to reduce $\Delta$ so as to realize a band inversion between \textit{s} and \textit{p} states and then increase $\lambda$ to open a nontrivial energy gap.

Additionally, interactions between different atomic sites also play an important role in determining the spectrum and localization of electronic states in quasicrystals \cite{PhysRevB.33.2184,PhysRevB.34.3904,PhysRevB.43.1378}. We further investigated the phase diagram with the increasing bonding strengths. As shown in Fig.~\ref{fig3}(b), the QL undergoes a topological phase transition from a NI to a QSH insulator at a critical value of about $d_0^c=0.89$. By further increasing the interaction, the system eventually enters a gapless phase which turns out to be a weak metal, as discussed later. Such a process can be understood by considering the evolution of states as following: starting from the atomic orbital limit, \textit{s} and \textit{p} states are initially separated by a trivial charge gap [Fig.~\ref{fig3}(c)]. By increasing the interaction one gradually enlarges the band width, therefore reduces and eventually closes the charge gap, realizing a \textit{s}-\textit{p} band inversion; the SOC effect then reopens an energy gap with nontrivial topology [Fig.~\ref{fig3}(d)]. Further increasing the interaction, to overcome the SOC gap, will drive the system into a gapless phase [Fig.~\ref{fig3}(e)]. Interestingly, in the QSH region a defect mode induced by periodic approximation of quasicrystal moves downwards into the energy gap gradually with the increasing $d_0$, while the whole gap remains to be topologically nontrivial with $B_s=1$. We also calculated phase diagrams of quasicrystal approximants with different sizes and found similar phase transitions in all approximants.\footnotemark[\value{footnote}] This implies that the topological phase transition as well as the QSH effect should appear in QLs in the thermodynamic limit of infinite lattice size.

Finally, we studied the localization of wavefunctions in the QL model. Whether the gapless state in the phase diagram is 
metallic or insulating depends on the localization of wavefunction around the Fermi level \cite{PhysRevB.34.3904,PhysRevB.43.1378}. For nonperiodic systems, we illustrate the localization of each state by its participation ratio \cite{PhysRevB.33.2184},
\begin{equation}
\gamma_n=\frac{(\sum_i^N|\langle i|\psi_n\rangle|^2)^2}{N\sum_i^N|\langle i|\psi_n\rangle|^4},
\end{equation}
where $|i\rangle$ is the \textit{i}-th local orbital. The participation ratio takes the value $1/N$ if the wavefunction is localized in a single orbital and unity if the wavefunction is extended uniformly over the whole system. The participation ratio of wavefunctions in QLs (Fig. 10 of Ref.~\footnotemark[\value{footnote}]) is less than 0.25 which is much smaller than that of extended wavefunctions in periodic crystals. However, the localization behavior is also different from disordered Anderson localization, where the state around the mobility gap tend to be more localized \cite{schreiber1985numerical, odagaki1986properties,de2013electronic}. This indicates that critical wavefunctions are induced by local structural topology of the QLs \cite{PhysRevB.51.15827, PhysRevB.34.5208}. The low participation ratio also suggests a weak metallic behavior in the electronic transport \cite{PhysRevB.35.1456,PhysRevB.38.10109, PhysRevB.43.8892, de2013electronic}. Our transport simulation gives a conductance about an order of magnitude smaller than that of pure periodic leads, confirming its weak metallic behavior \footnotemark[\value{footnote}].

\paragraph{Experimental feasibility.}
Our proposed atomic model of quasicrystalline QSH states is expected to be realized in surface-based 2D QLs. With the development of growth technique with atomic precision, 2D quasicrystals have been epitaxially grown on quasicrystalline substrates in the last two decades \cite{jenks1998quasicrystals, mcgrath2002quasicrystal, sharma2007quasicrystal, thiel2008quasicrystal,mcgrath2012memory, ledieu2014surfaces}. For example, single-element epitaxial quasicrystalline structures have been successfully grown by deposition of Si \cite{PhysRevB.73.012204}, Pb \cite{PhysRevB.77.073409,PhysRevB.82.085417, PhysRevB.79.165430, PhysRevB.79.245405,smerdon2008formation,sharma2013templated}, Sn \cite{PhysRevB.72.045428}, Sb \cite{PhysRevLett.89.156104}, Bi \cite{PhysRevB.78.075407}, Co \cite{smerdon2006adsorption}, and Cu atoms \cite{PhysRevLett.92.135507, PhysRevB.72.035420, PhysRevB.74.035429} on the fivefold (tenfold) surfaces of icosahedral (decagonal) quasicrystals which serve as template substrates. Moreover, aperiodic quasicrystalline phases can also be realized on crystalline surfaces \cite{forster2013quasicrystalline}. For example, atomically flat epitaxial Ag films with quasiperiodicity was synthesized on GaAs(110) surface \cite{ScienceAgGaAs}. Recently, Collins \textit{et al.} \cite{collins2017imaging} realized a synthetic QL by arranging carbon monoxide molecules on the surface of Cu(111) to form a Penrose tiling using scanning tunneling microscopy and atomic manipulation. We, therefore, expect that it is experimentally feasible to realize our theoretically proposed atomic model of QSH states on surface-based QLs.

\paragraph{Conclusion}
We have proposed the realization of QSH states in Penrose-type QLs. We characterize the topological nature by deriving a newly-defined topological invariant, the spin Bott index, in addition to conventional evidences including robust edge states and quantized conductances. Beyond the Penrose tiling, other QLs of different classes of local isomorphism should also be able to realize the QSH state \cite{senechal1996quasicrystals}. The essential band inversion is not limited to the $s$ and $p$ orbitals, and other types of band inversion mechanism \cite{zhang2013topological} such as \textit{p}-\textit{p} \cite{PhysRevB.93.035135}, \textit{p}-\textit{d} \cite{PhysRevLett.106.156808}, \textit{d}-\textit{d} \cite{si2016large}, and \textit{d}-\textit{f} band inversion \cite{PhysRevLett.104.106408}, are all feasible to achieve QSH states in quasicrystals. Our finding therefore extends the territory of topological materials beyond crystalline solids, to surface-based aperiodic systems with a range of choices of structural and element species. Our proposed approach is also applicable to other 2D QLs with different symmetries and to 3D quasicrystal structures \cite{PhysRevB.34.596,PhysRevB.34.617}, which may open additional exciting possibilities.

\begin{acknowledgments}
This work was supported by DOE-BES (Grant No. DE-FG02-04ER46148). The calculations were done on the CHPC at the University of Utah and DOE-NERSC.
\end{acknowledgments}


\begin{thebibliography}{88}%
\makeatletter
\providecommand \@ifxundefined [1]{%
 \@ifx{#1\undefined}
}%
\providecommand \@ifnum [1]{%
 \ifnum #1\expandafter \@firstoftwo
 \else \expandafter \@secondoftwo
 \fi
}%
\providecommand \@ifx [1]{%
 \ifx #1\expandafter \@firstoftwo
 \else \expandafter \@secondoftwo
 \fi
}%
\providecommand \natexlab [1]{#1}%
\providecommand \enquote  [1]{``#1''}%
\providecommand \bibnamefont  [1]{#1}%
\providecommand \bibfnamefont [1]{#1}%
\providecommand \citenamefont [1]{#1}%
\providecommand \href@noop [0]{\@secondoftwo}%
\providecommand \href [0]{\begingroup \@sanitize@url \@href}%
\providecommand \@href[1]{\@@startlink{#1}\@@href}%
\providecommand \@@href[1]{\endgroup#1\@@endlink}%
\providecommand \@sanitize@url [0]{\catcode `\\12\catcode `\$12\catcode
  `\&12\catcode `\#12\catcode `\^12\catcode `\_12\catcode `\%12\relax}%
\providecommand \@@startlink[1]{}%
\providecommand \@@endlink[0]{}%
\providecommand \url  [0]{\begingroup\@sanitize@url \@url }%
\providecommand \@url [1]{\endgroup\@href {#1}{\urlprefix }}%
\providecommand \urlprefix  [0]{URL }%
\providecommand \Eprint [0]{\href }%
\providecommand \doibase [0]{http://dx.doi.org/}%
\providecommand \selectlanguage [0]{\@gobble}%
\providecommand \bibinfo  [0]{\@secondoftwo}%
\providecommand \bibfield  [0]{\@secondoftwo}%
\providecommand \translation [1]{[#1]}%
\providecommand \BibitemOpen [0]{}%
\providecommand \bibitemStop [0]{}%
\providecommand \bibitemNoStop [0]{.\EOS\space}%
\providecommand \EOS [0]{\spacefactor3000\relax}%
\providecommand \BibitemShut  [1]{\csname bibitem#1\endcsname}%
\let\auto@bib@innerbib\@empty
\bibitem [{\citenamefont {Kane}\ and\ \citenamefont
  {Mele}(2005{\natexlab{a}})}]{PhysRevLett.95.226801}%
  \BibitemOpen
  \bibfield  {author} {\bibinfo {author} {\bibfnamefont {C.~L.}\ \bibnamefont
  {Kane}}\ and\ \bibinfo {author} {\bibfnamefont {E.~J.}\ \bibnamefont
  {Mele}},\ }\href {\doibase 10.1103/PhysRevLett.95.226801} {\bibfield
  {journal} {\bibinfo  {journal} {Phys. Rev. Lett.}\ }\textbf {\bibinfo
  {volume} {95}},\ \bibinfo {pages} {226801} (\bibinfo {year}
  {2005}{\natexlab{a}})}\BibitemShut {NoStop}%
\bibitem [{\citenamefont {Kane}\ and\ \citenamefont
  {Mele}(2005{\natexlab{b}})}]{PhysRevLett.95.146802}%
  \BibitemOpen
  \bibfield  {author} {\bibinfo {author} {\bibfnamefont {C.~L.}\ \bibnamefont
  {Kane}}\ and\ \bibinfo {author} {\bibfnamefont {E.~J.}\ \bibnamefont
  {Mele}},\ }\href {\doibase 10.1103/PhysRevLett.95.146802} {\bibfield
  {journal} {\bibinfo  {journal} {Phys. Rev. Lett.}\ }\textbf {\bibinfo
  {volume} {95}},\ \bibinfo {pages} {146802} (\bibinfo {year}
  {2005}{\natexlab{b}})}\BibitemShut {NoStop}%
\bibitem [{\citenamefont {Bernevig}\ and\ \citenamefont
  {Zhang}(2006)}]{PhysRevLett.96.106802}%
  \BibitemOpen
  \bibfield  {author} {\bibinfo {author} {\bibfnamefont {B.~A.}\ \bibnamefont
  {Bernevig}}\ and\ \bibinfo {author} {\bibfnamefont {S.-C.}\ \bibnamefont
  {Zhang}},\ }\href {\doibase 10.1103/PhysRevLett.96.106802} {\bibfield
  {journal} {\bibinfo  {journal} {Phys. Rev. Lett.}\ }\textbf {\bibinfo
  {volume} {96}},\ \bibinfo {pages} {106802} (\bibinfo {year}
  {2006})}\BibitemShut {NoStop}%
\bibitem [{\citenamefont {Bernevig}\ \emph {et~al.}(2006)\citenamefont
  {Bernevig}, \citenamefont {Hughes},\ and\ \citenamefont {Zhang}}]{BHZ}%
  \BibitemOpen
  \bibfield  {author} {\bibinfo {author} {\bibfnamefont {B.~A.}\ \bibnamefont
  {Bernevig}}, \bibinfo {author} {\bibfnamefont {T.~L.}\ \bibnamefont
  {Hughes}}, \ and\ \bibinfo {author} {\bibfnamefont {S.-C.}\ \bibnamefont
  {Zhang}},\ }\href {\doibase 10.1126/science.1133734} {\bibfield  {journal}
  {\bibinfo  {journal} {Science}\ }\textbf {\bibinfo {volume} {314}},\ \bibinfo
  {pages} {1757} (\bibinfo {year} {2006})}\BibitemShut {NoStop}%
\bibitem [{\citenamefont {K{\"o}nig}\ \emph {et~al.}(2007)\citenamefont
  {K{\"o}nig}, \citenamefont {Wiedmann}, \citenamefont {Br{\"u}ne},
  \citenamefont {Roth}, \citenamefont {Buhmann}, \citenamefont {Molenkamp},
  \citenamefont {Qi},\ and\ \citenamefont {Zhang}}]{konig}%
  \BibitemOpen
  \bibfield  {author} {\bibinfo {author} {\bibfnamefont {M.}~\bibnamefont
  {K{\"o}nig}}, \bibinfo {author} {\bibfnamefont {S.}~\bibnamefont {Wiedmann}},
  \bibinfo {author} {\bibfnamefont {C.}~\bibnamefont {Br{\"u}ne}}, \bibinfo
  {author} {\bibfnamefont {A.}~\bibnamefont {Roth}}, \bibinfo {author}
  {\bibfnamefont {H.}~\bibnamefont {Buhmann}}, \bibinfo {author} {\bibfnamefont
  {L.~W.}\ \bibnamefont {Molenkamp}}, \bibinfo {author} {\bibfnamefont {X.-L.}\
  \bibnamefont {Qi}}, \ and\ \bibinfo {author} {\bibfnamefont {S.-C.}\
  \bibnamefont {Zhang}},\ }\href@noop {} {\bibfield  {journal} {\bibinfo
  {journal} {Science}\ }\textbf {\bibinfo {volume} {318}},\ \bibinfo {pages}
  {766} (\bibinfo {year} {2007})}\BibitemShut {NoStop}%
\bibitem [{\citenamefont {Huang}\ \emph {et~al.}(2017)\citenamefont {Huang},
  \citenamefont {Xu}, \citenamefont {Wang},\ and\ \citenamefont
  {Duan}}]{huanghqCMS}%
  \BibitemOpen
  \bibfield  {author} {\bibinfo {author} {\bibfnamefont {H.}~\bibnamefont
  {Huang}}, \bibinfo {author} {\bibfnamefont {Y.}~\bibnamefont {Xu}}, \bibinfo
  {author} {\bibfnamefont {J.}~\bibnamefont {Wang}}, \ and\ \bibinfo {author}
  {\bibfnamefont {W.}~\bibnamefont {Duan}},\ }\href@noop {} {\bibfield
  {journal} {\bibinfo  {journal} {WIRES: Comp. Mol. Sci.}\ }\textbf {\bibinfo
  {volume} {7}} (\bibinfo {year} {2017})}\BibitemShut {NoStop}%
\bibitem [{\citenamefont {Wang}\ \emph {et~al.}(2017)\citenamefont {Wang},
  \citenamefont {Jin},\ and\ \citenamefont {Liu}}]{wang2017computational}%
  \BibitemOpen
  \bibfield  {author} {\bibinfo {author} {\bibfnamefont {Z.}~\bibnamefont
  {Wang}}, \bibinfo {author} {\bibfnamefont {K.-H.}\ \bibnamefont {Jin}}, \
  and\ \bibinfo {author} {\bibfnamefont {F.}~\bibnamefont {Liu}},\ }\href@noop
  {} {\bibfield  {journal} {\bibinfo  {journal} {WIRES: Comp. Mol. Sci.}\
  }\textbf {\bibinfo {volume} {7}} (\bibinfo {year} {2017})}\BibitemShut
  {NoStop}%
\bibitem [{\citenamefont {Zhou}\ \emph
  {et~al.}(2014{\natexlab{a}})\citenamefont {Zhou}, \citenamefont {Ming},
  \citenamefont {Liu}, \citenamefont {Wang}, \citenamefont {Li},\ and\
  \citenamefont {Liu}}]{zhou2014epitaxial}%
  \BibitemOpen
  \bibfield  {author} {\bibinfo {author} {\bibfnamefont {M.}~\bibnamefont
  {Zhou}}, \bibinfo {author} {\bibfnamefont {W.}~\bibnamefont {Ming}}, \bibinfo
  {author} {\bibfnamefont {Z.}~\bibnamefont {Liu}}, \bibinfo {author}
  {\bibfnamefont {Z.}~\bibnamefont {Wang}}, \bibinfo {author} {\bibfnamefont
  {P.}~\bibnamefont {Li}}, \ and\ \bibinfo {author} {\bibfnamefont
  {F.}~\bibnamefont {Liu}},\ }\href@noop {} {\bibfield  {journal} {\bibinfo
  {journal} {Proc. Natl. Acad. Sci.}\ }\textbf {\bibinfo {volume} {111}},\
  \bibinfo {pages} {14378} (\bibinfo {year} {2014}{\natexlab{a}})}\BibitemShut
  {NoStop}%
\bibitem [{\citenamefont {Zhou}\ \emph
  {et~al.}(2014{\natexlab{b}})\citenamefont {Zhou}, \citenamefont {Ming},
  \citenamefont {Liu}, \citenamefont {Wang}, \citenamefont {Yao},\ and\
  \citenamefont {Liu}}]{zhou2014formation}%
  \BibitemOpen
  \bibfield  {author} {\bibinfo {author} {\bibfnamefont {M.}~\bibnamefont
  {Zhou}}, \bibinfo {author} {\bibfnamefont {W.}~\bibnamefont {Ming}}, \bibinfo
  {author} {\bibfnamefont {Z.}~\bibnamefont {Liu}}, \bibinfo {author}
  {\bibfnamefont {Z.}~\bibnamefont {Wang}}, \bibinfo {author} {\bibfnamefont
  {Y.}~\bibnamefont {Yao}}, \ and\ \bibinfo {author} {\bibfnamefont
  {F.}~\bibnamefont {Liu}},\ }\href@noop {} {\bibfield  {journal} {\bibinfo
  {journal} {Sci. Rep.}\ }\textbf {\bibinfo {volume} {4}},\ \bibinfo {pages}
  {7102} (\bibinfo {year} {2014}{\natexlab{b}})}\BibitemShut {NoStop}%
\bibitem [{\citenamefont {Hsu}\ \emph {et~al.}(2015)\citenamefont {Hsu},
  \citenamefont {Huang}, \citenamefont {Chuang}, \citenamefont {Kuo},
  \citenamefont {Liu}, \citenamefont {Lin},\ and\ \citenamefont
  {Bansil}}]{hsu2015nontrivial}%
  \BibitemOpen
  \bibfield  {author} {\bibinfo {author} {\bibfnamefont {C.-H.}\ \bibnamefont
  {Hsu}}, \bibinfo {author} {\bibfnamefont {Z.-Q.}\ \bibnamefont {Huang}},
  \bibinfo {author} {\bibfnamefont {F.-C.}\ \bibnamefont {Chuang}}, \bibinfo
  {author} {\bibfnamefont {C.-C.}\ \bibnamefont {Kuo}}, \bibinfo {author}
  {\bibfnamefont {Y.-T.}\ \bibnamefont {Liu}}, \bibinfo {author} {\bibfnamefont
  {H.}~\bibnamefont {Lin}}, \ and\ \bibinfo {author} {\bibfnamefont
  {A.}~\bibnamefont {Bansil}},\ }\href@noop {} {\bibfield  {journal} {\bibinfo
  {journal} {New J. Phys.}\ }\textbf {\bibinfo {volume} {17}},\ \bibinfo
  {pages} {025005} (\bibinfo {year} {2015})}\BibitemShut {NoStop}%
\bibitem [{\citenamefont {Wang}\ \emph {et~al.}(2016)\citenamefont {Wang},
  \citenamefont {Jin},\ and\ \citenamefont {Liu}}]{wang2016quantum}%
  \BibitemOpen
  \bibfield  {author} {\bibinfo {author} {\bibfnamefont {Z.}~\bibnamefont
  {Wang}}, \bibinfo {author} {\bibfnamefont {K.-H.}\ \bibnamefont {Jin}}, \
  and\ \bibinfo {author} {\bibfnamefont {F.}~\bibnamefont {Liu}},\ }\href@noop
  {} {\bibfield  {journal} {\bibinfo  {journal} {Nature Commun.}\ }\textbf
  {\bibinfo {volume} {7}},\ \bibinfo {pages} {12746} (\bibinfo {year}
  {2016})}\BibitemShut {NoStop}%
\bibitem [{\citenamefont {Reis}\ \emph {et~al.}(2017)\citenamefont {Reis},
  \citenamefont {Li}, \citenamefont {Dudy}, \citenamefont {Bauernfeind},
  \citenamefont {Glass}, \citenamefont {Hanke}, \citenamefont {Thomale},
  \citenamefont {Sch{\"a}fer},\ and\ \citenamefont
  {Claessen}}]{reis2017bismuthene}%
  \BibitemOpen
  \bibfield  {author} {\bibinfo {author} {\bibfnamefont {F.}~\bibnamefont
  {Reis}}, \bibinfo {author} {\bibfnamefont {G.}~\bibnamefont {Li}}, \bibinfo
  {author} {\bibfnamefont {L.}~\bibnamefont {Dudy}}, \bibinfo {author}
  {\bibfnamefont {M.}~\bibnamefont {Bauernfeind}}, \bibinfo {author}
  {\bibfnamefont {S.}~\bibnamefont {Glass}}, \bibinfo {author} {\bibfnamefont
  {W.}~\bibnamefont {Hanke}}, \bibinfo {author} {\bibfnamefont
  {R.}~\bibnamefont {Thomale}}, \bibinfo {author} {\bibfnamefont
  {J.}~\bibnamefont {Sch{\"a}fer}}, \ and\ \bibinfo {author} {\bibfnamefont
  {R.}~\bibnamefont {Claessen}},\ }\href@noop {} {\bibfield  {journal}
  {\bibinfo  {journal} {Science}\ }\textbf {\bibinfo {volume} {357}},\ \bibinfo
  {pages} {287} (\bibinfo {year} {2017})}\BibitemShut {NoStop}%
\bibitem [{\citenamefont {Fu}\ and\ \citenamefont
  {Kane}(2006)}]{PhysRevB.74.195312}%
  \BibitemOpen
  \bibfield  {author} {\bibinfo {author} {\bibfnamefont {L.}~\bibnamefont
  {Fu}}\ and\ \bibinfo {author} {\bibfnamefont {C.~L.}\ \bibnamefont {Kane}},\
  }\href {\doibase 10.1103/PhysRevB.74.195312} {\bibfield  {journal} {\bibinfo
  {journal} {Phys. Rev. B}\ }\textbf {\bibinfo {volume} {74}},\ \bibinfo
  {pages} {195312} (\bibinfo {year} {2006})}\BibitemShut {NoStop}%
\bibitem [{\citenamefont {Shechtman}\ \emph {et~al.}(1984)\citenamefont
  {Shechtman}, \citenamefont {Blech}, \citenamefont {Gratias},\ and\
  \citenamefont {Cahn}}]{PhysRevLett.53.1951}%
  \BibitemOpen
  \bibfield  {author} {\bibinfo {author} {\bibfnamefont {D.}~\bibnamefont
  {Shechtman}}, \bibinfo {author} {\bibfnamefont {I.}~\bibnamefont {Blech}},
  \bibinfo {author} {\bibfnamefont {D.}~\bibnamefont {Gratias}}, \ and\
  \bibinfo {author} {\bibfnamefont {J.~W.}\ \bibnamefont {Cahn}},\ }\href
  {\doibase 10.1103/PhysRevLett.53.1951} {\bibfield  {journal} {\bibinfo
  {journal} {Phys. Rev. Lett.}\ }\textbf {\bibinfo {volume} {53}},\ \bibinfo
  {pages} {1951} (\bibinfo {year} {1984})}\BibitemShut {NoStop}%
\bibitem [{\citenamefont {Dubois}\ \emph {et~al.}(1991)\citenamefont {Dubois},
  \citenamefont {Kang},\ and\ \citenamefont
  {Von~Stebut}}]{dubois1991quasicrystalline}%
  \BibitemOpen
  \bibfield  {author} {\bibinfo {author} {\bibfnamefont {J.}~\bibnamefont
  {Dubois}}, \bibinfo {author} {\bibfnamefont {S.}~\bibnamefont {Kang}}, \ and\
  \bibinfo {author} {\bibfnamefont {J.}~\bibnamefont {Von~Stebut}},\
  }\href@noop {} {\bibfield  {journal} {\bibinfo  {journal} {J. Mater. Sci.
  Lett.}\ }\textbf {\bibinfo {volume} {10}},\ \bibinfo {pages} {537} (\bibinfo
  {year} {1991})}\BibitemShut {NoStop}%
\bibitem [{\citenamefont {Sutherland}(1986{\natexlab{a}})}]{PhysRevB.34.3904}%
  \BibitemOpen
  \bibfield  {author} {\bibinfo {author} {\bibfnamefont {B.}~\bibnamefont
  {Sutherland}},\ }\href {\doibase 10.1103/PhysRevB.34.3904} {\bibfield
  {journal} {\bibinfo  {journal} {Phys. Rev. B}\ }\textbf {\bibinfo {volume}
  {34}},\ \bibinfo {pages} {3904} (\bibinfo {year}
  {1986}{\natexlab{a}})}\BibitemShut {NoStop}%
\bibitem [{\citenamefont {Fujiwara}\ \emph {et~al.}(1988)\citenamefont
  {Fujiwara}, \citenamefont {Arai}, \citenamefont {Tokihiro},\ and\
  \citenamefont {Kohmoto}}]{PhysRevB.37.2797}%
  \BibitemOpen
  \bibfield  {author} {\bibinfo {author} {\bibfnamefont {T.}~\bibnamefont
  {Fujiwara}}, \bibinfo {author} {\bibfnamefont {M.}~\bibnamefont {Arai}},
  \bibinfo {author} {\bibfnamefont {T.}~\bibnamefont {Tokihiro}}, \ and\
  \bibinfo {author} {\bibfnamefont {M.}~\bibnamefont {Kohmoto}},\ }\href
  {\doibase 10.1103/PhysRevB.37.2797} {\bibfield  {journal} {\bibinfo
  {journal} {Phys. Rev. B}\ }\textbf {\bibinfo {volume} {37}},\ \bibinfo
  {pages} {2797} (\bibinfo {year} {1988})}\BibitemShut {NoStop}%
\bibitem [{\citenamefont {Tsunetsugu}\ \emph {et~al.}(1991)\citenamefont
  {Tsunetsugu}, \citenamefont {Fujiwara}, \citenamefont {Ueda},\ and\
  \citenamefont {Tokihiro}}]{PhysRevB.43.8879}%
  \BibitemOpen
  \bibfield  {author} {\bibinfo {author} {\bibfnamefont {H.}~\bibnamefont
  {Tsunetsugu}}, \bibinfo {author} {\bibfnamefont {T.}~\bibnamefont
  {Fujiwara}}, \bibinfo {author} {\bibfnamefont {K.}~\bibnamefont {Ueda}}, \
  and\ \bibinfo {author} {\bibfnamefont {T.}~\bibnamefont {Tokihiro}},\ }\href
  {\doibase 10.1103/PhysRevB.43.8879} {\bibfield  {journal} {\bibinfo
  {journal} {Phys. Rev. B}\ }\textbf {\bibinfo {volume} {43}},\ \bibinfo
  {pages} {8879} (\bibinfo {year} {1991})}\BibitemShut {NoStop}%
\bibitem [{\citenamefont {Tsunetsugu}\ and\ \citenamefont
  {Ueda}(1991)}]{PhysRevB.43.8892}%
  \BibitemOpen
  \bibfield  {author} {\bibinfo {author} {\bibfnamefont {H.}~\bibnamefont
  {Tsunetsugu}}\ and\ \bibinfo {author} {\bibfnamefont {K.}~\bibnamefont
  {Ueda}},\ }\href {\doibase 10.1103/PhysRevB.43.8892} {\bibfield  {journal}
  {\bibinfo  {journal} {Phys. Rev. B}\ }\textbf {\bibinfo {volume} {43}},\
  \bibinfo {pages} {8892} (\bibinfo {year} {1991})}\BibitemShut {NoStop}%
\bibitem [{\citenamefont {Jenks}\ and\ \citenamefont
  {Thiel}(1998)}]{jenks1998quasicrystals}%
  \BibitemOpen
  \bibfield  {author} {\bibinfo {author} {\bibfnamefont {C.~J.}\ \bibnamefont
  {Jenks}}\ and\ \bibinfo {author} {\bibfnamefont {P.~A.}\ \bibnamefont
  {Thiel}},\ }\href@noop {} {\bibfield  {journal} {\bibinfo  {journal}
  {Langmuir}\ }\textbf {\bibinfo {volume} {14}},\ \bibinfo {pages} {1392}
  (\bibinfo {year} {1998})}\BibitemShut {NoStop}%
\bibitem [{\citenamefont {McGrath}\ \emph {et~al.}(2002)\citenamefont
  {McGrath}, \citenamefont {Ledieu}, \citenamefont {Cox},\ and\ \citenamefont
  {Diehl}}]{mcgrath2002quasicrystal}%
  \BibitemOpen
  \bibfield  {author} {\bibinfo {author} {\bibfnamefont {R.}~\bibnamefont
  {McGrath}}, \bibinfo {author} {\bibfnamefont {J.}~\bibnamefont {Ledieu}},
  \bibinfo {author} {\bibfnamefont {E.~J.}\ \bibnamefont {Cox}}, \ and\
  \bibinfo {author} {\bibfnamefont {R.~D.}\ \bibnamefont {Diehl}},\ }\href@noop
  {} {\bibfield  {journal} {\bibinfo  {journal} {J. Phys. Condens. Matter}\
  }\textbf {\bibinfo {volume} {14}},\ \bibinfo {pages} {R119} (\bibinfo {year}
  {2002})}\BibitemShut {NoStop}%
\bibitem [{\citenamefont {Sharma}\ \emph {et~al.}(2007)\citenamefont {Sharma},
  \citenamefont {Shimoda},\ and\ \citenamefont
  {Tsai}}]{sharma2007quasicrystal}%
  \BibitemOpen
  \bibfield  {author} {\bibinfo {author} {\bibfnamefont {H.}~\bibnamefont
  {Sharma}}, \bibinfo {author} {\bibfnamefont {M.}~\bibnamefont {Shimoda}}, \
  and\ \bibinfo {author} {\bibfnamefont {A.}~\bibnamefont {Tsai}},\ }\href@noop
  {} {\bibfield  {journal} {\bibinfo  {journal} {Adv. Phys.}\ }\textbf
  {\bibinfo {volume} {56}},\ \bibinfo {pages} {403} (\bibinfo {year}
  {2007})}\BibitemShut {NoStop}%
\bibitem [{\citenamefont {Thiel}(2008)}]{thiel2008quasicrystal}%
  \BibitemOpen
  \bibfield  {author} {\bibinfo {author} {\bibfnamefont {P.~A.}\ \bibnamefont
  {Thiel}},\ }\href@noop {} {\bibfield  {journal} {\bibinfo  {journal} {Annu.
  Rev. Phys. Chem.}\ }\textbf {\bibinfo {volume} {59}},\ \bibinfo {pages} {129}
  (\bibinfo {year} {2008})}\BibitemShut {NoStop}%
\bibitem [{\citenamefont {McGrath}\ \emph {et~al.}(2012)\citenamefont
  {McGrath}, \citenamefont {Sharma}, \citenamefont {Smerdon},\ and\
  \citenamefont {Ledieu}}]{mcgrath2012memory}%
  \BibitemOpen
  \bibfield  {author} {\bibinfo {author} {\bibfnamefont {R.}~\bibnamefont
  {McGrath}}, \bibinfo {author} {\bibfnamefont {H.}~\bibnamefont {Sharma}},
  \bibinfo {author} {\bibfnamefont {J.}~\bibnamefont {Smerdon}}, \ and\
  \bibinfo {author} {\bibfnamefont {J.}~\bibnamefont {Ledieu}},\ }\href@noop {}
  {\bibfield  {journal} {\bibinfo  {journal} {Phil. Trans. R. Soc. A}\ }\textbf
  {\bibinfo {volume} {370}},\ \bibinfo {pages} {2930} (\bibinfo {year}
  {2012})}\BibitemShut {NoStop}%
\bibitem [{\citenamefont {Ledieu}\ and\ \citenamefont
  {Fourn{\'e}e}(2014)}]{ledieu2014surfaces}%
  \BibitemOpen
  \bibfield  {author} {\bibinfo {author} {\bibfnamefont {J.}~\bibnamefont
  {Ledieu}}\ and\ \bibinfo {author} {\bibfnamefont {V.}~\bibnamefont
  {Fourn{\'e}e}},\ }\href@noop {} {\bibfield  {journal} {\bibinfo  {journal}
  {Comptes Rendus Physique}\ }\textbf {\bibinfo {volume} {15}},\ \bibinfo
  {pages} {48} (\bibinfo {year} {2014})}\BibitemShut {NoStop}%
\bibitem [{\citenamefont {Bandres}\ \emph {et~al.}(2016)\citenamefont
  {Bandres}, \citenamefont {Rechtsman},\ and\ \citenamefont
  {Segev}}]{PhysRevX.6.011016}%
  \BibitemOpen
  \bibfield  {author} {\bibinfo {author} {\bibfnamefont {M.~A.}\ \bibnamefont
  {Bandres}}, \bibinfo {author} {\bibfnamefont {M.~C.}\ \bibnamefont
  {Rechtsman}}, \ and\ \bibinfo {author} {\bibfnamefont {M.}~\bibnamefont
  {Segev}},\ }\href {\doibase 10.1103/PhysRevX.6.011016} {\bibfield  {journal}
  {\bibinfo  {journal} {Phys. Rev. X}\ }\textbf {\bibinfo {volume} {6}},\
  \bibinfo {pages} {011016} (\bibinfo {year} {2016})}\BibitemShut {NoStop}%
\bibitem [{\citenamefont {Mitchell}\ \emph {et~al.}(2018)\citenamefont
  {Mitchell}, \citenamefont {Nash}, \citenamefont {Hexner}, \citenamefont
  {Turner},\ and\ \citenamefont {Irvine}}]{mitchell2018amorphous}%
  \BibitemOpen
  \bibfield  {author} {\bibinfo {author} {\bibfnamefont {N.~P.}\ \bibnamefont
  {Mitchell}}, \bibinfo {author} {\bibfnamefont {L.~M.}\ \bibnamefont {Nash}},
  \bibinfo {author} {\bibfnamefont {D.}~\bibnamefont {Hexner}}, \bibinfo
  {author} {\bibfnamefont {A.~M.}\ \bibnamefont {Turner}}, \ and\ \bibinfo
  {author} {\bibfnamefont {W.~T.}\ \bibnamefont {Irvine}},\ }\href@noop {}
  {\bibfield  {journal} {\bibinfo  {journal} {Nat. Phys.}\ ,\ \bibinfo {pages}
  {1}} (\bibinfo {year} {2018})}\BibitemShut {NoStop}%
\bibitem [{\citenamefont {Agarwala}\ and\ \citenamefont
  {Shenoy}(2017)}]{PhysRevLett.118.236402}%
  \BibitemOpen
  \bibfield  {author} {\bibinfo {author} {\bibfnamefont {A.}~\bibnamefont
  {Agarwala}}\ and\ \bibinfo {author} {\bibfnamefont {V.~B.}\ \bibnamefont
  {Shenoy}},\ }\href {\doibase 10.1103/PhysRevLett.118.236402} {\bibfield
  {journal} {\bibinfo  {journal} {Phys. Rev. Lett.}\ }\textbf {\bibinfo
  {volume} {118}},\ \bibinfo {pages} {236402} (\bibinfo {year}
  {2017})}\BibitemShut {NoStop}%
\bibitem [{\citenamefont {Penrose}(1974)}]{penrose1974role}%
  \BibitemOpen
  \bibfield  {author} {\bibinfo {author} {\bibfnamefont {R.}~\bibnamefont
  {Penrose}},\ }\href@noop {} {\bibfield  {journal} {\bibinfo  {journal} {Bull.
  Inst. Math. Appl.}\ }\textbf {\bibinfo {volume} {10}},\ \bibinfo {pages}
  {266} (\bibinfo {year} {1974})}\BibitemShut {NoStop}%
\bibitem [{\citenamefont {Tsunetsugu}\ \emph {et~al.}(1986)\citenamefont
  {Tsunetsugu}, \citenamefont {Fujiwara}, \citenamefont {Ueda},\ and\
  \citenamefont {Tokihiro}}]{tsunetsugu1986eigenstates}%
  \BibitemOpen
  \bibfield  {author} {\bibinfo {author} {\bibfnamefont {H.}~\bibnamefont
  {Tsunetsugu}}, \bibinfo {author} {\bibfnamefont {T.}~\bibnamefont
  {Fujiwara}}, \bibinfo {author} {\bibfnamefont {K.}~\bibnamefont {Ueda}}, \
  and\ \bibinfo {author} {\bibfnamefont {T.}~\bibnamefont {Tokihiro}},\
  }\href@noop {} {\bibfield  {journal} {\bibinfo  {journal} {J. Phys. Soc.
  Jpn.}\ }\textbf {\bibinfo {volume} {55}},\ \bibinfo {pages} {1420} (\bibinfo
  {year} {1986})}\BibitemShut {NoStop}%
\bibitem [{\citenamefont {Entin-Wohlman}\ \emph {et~al.}(1988)\citenamefont
  {Entin-Wohlman}, \citenamefont {Kl{\'e}man},\ and\ \citenamefont
  {Pavlovitch}}]{entin1988penrose}%
  \BibitemOpen
  \bibfield  {author} {\bibinfo {author} {\bibfnamefont {O.}~\bibnamefont
  {Entin-Wohlman}}, \bibinfo {author} {\bibfnamefont {M.}~\bibnamefont
  {Kl{\'e}man}}, \ and\ \bibinfo {author} {\bibfnamefont {A.}~\bibnamefont
  {Pavlovitch}},\ }\href@noop {} {\bibfield  {journal} {\bibinfo  {journal}
  {Journal de Physique}\ }\textbf {\bibinfo {volume} {49}},\ \bibinfo {pages}
  {587} (\bibinfo {year} {1988})}\BibitemShut {NoStop}%
\bibitem [{\citenamefont {Slater}\ and\ \citenamefont
  {Koster}(1954)}]{SlaterKoster}%
  \BibitemOpen
  \bibfield  {author} {\bibinfo {author} {\bibfnamefont {J.~C.}\ \bibnamefont
  {Slater}}\ and\ \bibinfo {author} {\bibfnamefont {G.~F.}\ \bibnamefont
  {Koster}},\ }\href {\doibase 10.1103/PhysRev.94.1498} {\bibfield  {journal}
  {\bibinfo  {journal} {Phys. Rev.}\ }\textbf {\bibinfo {volume} {94}},\
  \bibinfo {pages} {1498} (\bibinfo {year} {1954})}\BibitemShut {NoStop}%
\bibitem [{\citenamefont {Harrison}(2012)}]{harrison2012electronic}%
  \BibitemOpen
  \bibfield  {author} {\bibinfo {author} {\bibfnamefont {W.~A.}\ \bibnamefont
  {Harrison}},\ }\href@noop {} {\emph {\bibinfo {title} {Electronic structure
  and the properties of solids: the physics of the chemical bond}}}\ (\bibinfo
  {publisher} {Courier Corporation},\ \bibinfo {year} {2012})\BibitemShut
  {NoStop}%
\bibitem [{Note1()}]{Note1}%
  \BibitemOpen
  \bibinfo {note} {\label {fn} See the joint publication: H. Huang and F. Liu,
  Phys. Rev. B xx,xxx (2018), for more details.}\BibitemShut {Stop}%
\bibitem [{\citenamefont {Datta}(1997)}]{datta1997electronic}%
  \BibitemOpen
  \bibfield  {author} {\bibinfo {author} {\bibfnamefont {S.}~\bibnamefont
  {Datta}},\ }\href@noop {} {\emph {\bibinfo {title} {Electronic transport in
  mesoscopic systems}}}\ (\bibinfo  {publisher} {Cambridge university press},\
  \bibinfo {year} {1997})\BibitemShut {NoStop}%
\bibitem [{\citenamefont {B\"uttiker}(1988)}]{PhysRevB.38.9375}%
  \BibitemOpen
  \bibfield  {author} {\bibinfo {author} {\bibfnamefont {M.}~\bibnamefont
  {B\"uttiker}},\ }\href {\doibase 10.1103/PhysRevB.38.9375} {\bibfield
  {journal} {\bibinfo  {journal} {Phys. Rev. B}\ }\textbf {\bibinfo {volume}
  {38}},\ \bibinfo {pages} {9375} (\bibinfo {year} {1988})}\BibitemShut
  {NoStop}%
\bibitem [{\citenamefont {Huang}\ \emph
  {et~al.}(2015{\natexlab{a}})\citenamefont {Huang}, \citenamefont {Wang},
  \citenamefont {Luo}, \citenamefont {Liu}, \citenamefont {L\"u}, \citenamefont
  {Wu},\ and\ \citenamefont {Duan}}]{huanghqInterface}%
  \BibitemOpen
  \bibfield  {author} {\bibinfo {author} {\bibfnamefont {H.}~\bibnamefont
  {Huang}}, \bibinfo {author} {\bibfnamefont {Z.}~\bibnamefont {Wang}},
  \bibinfo {author} {\bibfnamefont {N.}~\bibnamefont {Luo}}, \bibinfo {author}
  {\bibfnamefont {Z.}~\bibnamefont {Liu}}, \bibinfo {author} {\bibfnamefont
  {R.}~\bibnamefont {L\"u}}, \bibinfo {author} {\bibfnamefont {J.}~\bibnamefont
  {Wu}}, \ and\ \bibinfo {author} {\bibfnamefont {W.}~\bibnamefont {Duan}},\
  }\href {\doibase 10.1103/PhysRevB.92.075138} {\bibfield  {journal} {\bibinfo
  {journal} {Phys. Rev. B}\ }\textbf {\bibinfo {volume} {92}},\ \bibinfo
  {pages} {075138} (\bibinfo {year} {2015}{\natexlab{a}})}\BibitemShut
  {NoStop}%
\bibitem [{\citenamefont {Haldane}(1988)}]{Haldane}%
  \BibitemOpen
  \bibfield  {author} {\bibinfo {author} {\bibfnamefont {F.~D.~M.}\
  \bibnamefont {Haldane}},\ }\href {\doibase 10.1103/PhysRevLett.61.2015}
  {\bibfield  {journal} {\bibinfo  {journal} {Phys. Rev. Lett.}\ }\textbf
  {\bibinfo {volume} {61}},\ \bibinfo {pages} {2015} (\bibinfo {year}
  {1988})}\BibitemShut {NoStop}%
\bibitem [{\citenamefont {Chang}\ \emph {et~al.}(2013)\citenamefont {Chang},
  \citenamefont {Zhang}, \citenamefont {Feng}, \citenamefont {Shen},
  \citenamefont {Zhang}, \citenamefont {Guo}, \citenamefont {Li}, \citenamefont
  {Ou}, \citenamefont {Wei}, \citenamefont {Wang} \emph
  {et~al.}}]{chang2013experimental}%
  \BibitemOpen
  \bibfield  {author} {\bibinfo {author} {\bibfnamefont {C.-Z.}\ \bibnamefont
  {Chang}}, \bibinfo {author} {\bibfnamefont {J.}~\bibnamefont {Zhang}},
  \bibinfo {author} {\bibfnamefont {X.}~\bibnamefont {Feng}}, \bibinfo {author}
  {\bibfnamefont {J.}~\bibnamefont {Shen}}, \bibinfo {author} {\bibfnamefont
  {Z.}~\bibnamefont {Zhang}}, \bibinfo {author} {\bibfnamefont
  {M.}~\bibnamefont {Guo}}, \bibinfo {author} {\bibfnamefont {K.}~\bibnamefont
  {Li}}, \bibinfo {author} {\bibfnamefont {Y.}~\bibnamefont {Ou}}, \bibinfo
  {author} {\bibfnamefont {P.}~\bibnamefont {Wei}}, \bibinfo {author}
  {\bibfnamefont {L.-L.}\ \bibnamefont {Wang}},  \emph {et~al.},\ }\href@noop
  {} {\bibfield  {journal} {\bibinfo  {journal} {Science}\ }\textbf {\bibinfo
  {volume} {340}},\ \bibinfo {pages} {167} (\bibinfo {year}
  {2013})}\BibitemShut {NoStop}%
\bibitem [{\citenamefont {Huang}\ \emph
  {et~al.}(2015{\natexlab{b}})\citenamefont {Huang}, \citenamefont {Liu},
  \citenamefont {Zhang}, \citenamefont {Duan},\ and\ \citenamefont
  {Vanderbilt}}]{huanghqSemiDirac}%
  \BibitemOpen
  \bibfield  {author} {\bibinfo {author} {\bibfnamefont {H.}~\bibnamefont
  {Huang}}, \bibinfo {author} {\bibfnamefont {Z.}~\bibnamefont {Liu}}, \bibinfo
  {author} {\bibfnamefont {H.}~\bibnamefont {Zhang}}, \bibinfo {author}
  {\bibfnamefont {W.}~\bibnamefont {Duan}}, \ and\ \bibinfo {author}
  {\bibfnamefont {D.}~\bibnamefont {Vanderbilt}},\ }\href {\doibase
  10.1103/PhysRevB.92.161115} {\bibfield  {journal} {\bibinfo  {journal} {Phys.
  Rev. B}\ }\textbf {\bibinfo {volume} {92}},\ \bibinfo {pages} {161115}
  (\bibinfo {year} {2015}{\natexlab{b}})}\BibitemShut {NoStop}%
\bibitem [{\citenamefont {Toniolo}(2017)}]{toniolo2017equivalence}%
  \BibitemOpen
  \bibfield  {author} {\bibinfo {author} {\bibfnamefont {D.}~\bibnamefont
  {Toniolo}},\ }\href@noop {} {\bibfield  {journal} {\bibinfo  {journal}
  {arXiv:1708.05912}\ } (\bibinfo {year} {2017})}\BibitemShut {NoStop}%
\bibitem [{\citenamefont {Loring}\ and\ \citenamefont
  {Hastings}(2011)}]{loring2011disordered}%
  \BibitemOpen
  \bibfield  {author} {\bibinfo {author} {\bibfnamefont {T.~A.}\ \bibnamefont
  {Loring}}\ and\ \bibinfo {author} {\bibfnamefont {M.~B.}\ \bibnamefont
  {Hastings}},\ }\href@noop {} {\bibfield  {journal} {\bibinfo  {journal} {EPL
  (Europhysics Letters)}\ }\textbf {\bibinfo {volume} {92}},\ \bibinfo {pages}
  {67004} (\bibinfo {year} {2011})}\BibitemShut {NoStop}%
\bibitem [{\citenamefont {Hastings}\ and\ \citenamefont
  {Loring}(2011)}]{hastings2011topological}%
  \BibitemOpen
  \bibfield  {author} {\bibinfo {author} {\bibfnamefont {M.~B.}\ \bibnamefont
  {Hastings}}\ and\ \bibinfo {author} {\bibfnamefont {T.~A.}\ \bibnamefont
  {Loring}},\ }\href@noop {} {\bibfield  {journal} {\bibinfo  {journal} {Ann.
  Phys.}\ }\textbf {\bibinfo {volume} {326}},\ \bibinfo {pages} {1699}
  (\bibinfo {year} {2011})}\BibitemShut {NoStop}%
\bibitem [{\citenamefont {Loring}(2015)}]{loring2015k}%
  \BibitemOpen
  \bibfield  {author} {\bibinfo {author} {\bibfnamefont {T.~A.}\ \bibnamefont
  {Loring}},\ }\href@noop {} {\bibfield  {journal} {\bibinfo  {journal} {Ann.
  Phys.}\ }\textbf {\bibinfo {volume} {356}},\ \bibinfo {pages} {383} (\bibinfo
  {year} {2015})}\BibitemShut {NoStop}%
\bibitem [{\citenamefont {Sheng}\ \emph {et~al.}(2006)\citenamefont {Sheng},
  \citenamefont {Weng}, \citenamefont {Sheng},\ and\ \citenamefont
  {Haldane}}]{PhysRevLett.97.036808}%
  \BibitemOpen
  \bibfield  {author} {\bibinfo {author} {\bibfnamefont {D.~N.}\ \bibnamefont
  {Sheng}}, \bibinfo {author} {\bibfnamefont {Z.~Y.}\ \bibnamefont {Weng}},
  \bibinfo {author} {\bibfnamefont {L.}~\bibnamefont {Sheng}}, \ and\ \bibinfo
  {author} {\bibfnamefont {F.~D.~M.}\ \bibnamefont {Haldane}},\ }\href
  {\doibase 10.1103/PhysRevLett.97.036808} {\bibfield  {journal} {\bibinfo
  {journal} {Phys. Rev. Lett.}\ }\textbf {\bibinfo {volume} {97}},\ \bibinfo
  {pages} {036808} (\bibinfo {year} {2006})}\BibitemShut {NoStop}%
\bibitem [{\citenamefont {Fukui}\ and\ \citenamefont
  {Hatsugai}(2007)}]{PhysRevB.75.121403}%
  \BibitemOpen
  \bibfield  {author} {\bibinfo {author} {\bibfnamefont {T.}~\bibnamefont
  {Fukui}}\ and\ \bibinfo {author} {\bibfnamefont {Y.}~\bibnamefont
  {Hatsugai}},\ }\href {\doibase 10.1103/PhysRevB.75.121403} {\bibfield
  {journal} {\bibinfo  {journal} {Phys. Rev. B}\ }\textbf {\bibinfo {volume}
  {75}},\ \bibinfo {pages} {121403} (\bibinfo {year} {2007})}\BibitemShut
  {NoStop}%
\bibitem [{\citenamefont {Prodan}(2009)}]{PhysRevB.80.125327}%
  \BibitemOpen
  \bibfield  {author} {\bibinfo {author} {\bibfnamefont {E.}~\bibnamefont
  {Prodan}},\ }\href {\doibase 10.1103/PhysRevB.80.125327} {\bibfield
  {journal} {\bibinfo  {journal} {Phys. Rev. B}\ }\textbf {\bibinfo {volume}
  {80}},\ \bibinfo {pages} {125327} (\bibinfo {year} {2009})}\BibitemShut
  {NoStop}%
\bibitem [{\citenamefont {Prodan}(2010)}]{prodan2010non}%
  \BibitemOpen
  \bibfield  {author} {\bibinfo {author} {\bibfnamefont {E.}~\bibnamefont
  {Prodan}},\ }\href@noop {} {\bibfield  {journal} {\bibinfo  {journal} {New J.
  Phys.}\ }\textbf {\bibinfo {volume} {12}},\ \bibinfo {pages} {065003}
  (\bibinfo {year} {2010})}\BibitemShut {NoStop}%
\bibitem [{\citenamefont {Prodan}(2011)}]{prodan2011disordered}%
  \BibitemOpen
  \bibfield  {author} {\bibinfo {author} {\bibfnamefont {E.}~\bibnamefont
  {Prodan}},\ }\href@noop {} {\bibfield  {journal} {\bibinfo  {journal} {J.
  Phys A: Math. Theor.}\ }\textbf {\bibinfo {volume} {44}},\ \bibinfo {pages}
  {113001} (\bibinfo {year} {2011})}\BibitemShut {NoStop}%
\bibitem [{\citenamefont {Bellissard}\ \emph {et~al.}(1994)\citenamefont
  {Bellissard}, \citenamefont {van Elst},\ and\ \citenamefont
  {Schulz-Baldes}}]{bellissard1994noncommutative}%
  \BibitemOpen
  \bibfield  {author} {\bibinfo {author} {\bibfnamefont {J.}~\bibnamefont
  {Bellissard}}, \bibinfo {author} {\bibfnamefont {A.}~\bibnamefont {van
  Elst}}, \ and\ \bibinfo {author} {\bibfnamefont {H.}~\bibnamefont
  {Schulz-Baldes}},\ }\href@noop {} {\bibfield  {journal} {\bibinfo  {journal}
  {J. Math. Phys.}\ }\textbf {\bibinfo {volume} {35}},\ \bibinfo {pages} {5373}
  (\bibinfo {year} {1994})}\BibitemShut {NoStop}%
\bibitem [{\citenamefont {Hastings}\ and\ \citenamefont
  {Loring}(2010)}]{hastings2010almost}%
  \BibitemOpen
  \bibfield  {author} {\bibinfo {author} {\bibfnamefont {M.~B.}\ \bibnamefont
  {Hastings}}\ and\ \bibinfo {author} {\bibfnamefont {T.~A.}\ \bibnamefont
  {Loring}},\ }\href@noop {} {\bibfield  {journal} {\bibinfo  {journal} {J.
  Math. Phys.}\ }\textbf {\bibinfo {volume} {51}},\ \bibinfo {pages} {015214}
  (\bibinfo {year} {2010})}\BibitemShut {NoStop}%
\bibitem [{\citenamefont {Exel}\ and\ \citenamefont
  {Loring}(1991)}]{exel1991invariants}%
  \BibitemOpen
  \bibfield  {author} {\bibinfo {author} {\bibfnamefont {R.}~\bibnamefont
  {Exel}}\ and\ \bibinfo {author} {\bibfnamefont {T.~A.}\ \bibnamefont
  {Loring}},\ }\href@noop {} {\bibfield  {journal} {\bibinfo  {journal} {J.
  Funct. Anal.}\ }\textbf {\bibinfo {volume} {95}},\ \bibinfo {pages} {364}
  (\bibinfo {year} {1991})}\BibitemShut {NoStop}%
\bibitem [{\citenamefont {Katsura}\ and\ \citenamefont
  {Koma}(2016)}]{katsura2016Z2}%
  \BibitemOpen
  \bibfield  {author} {\bibinfo {author} {\bibfnamefont {H.}~\bibnamefont
  {Katsura}}\ and\ \bibinfo {author} {\bibfnamefont {T.}~\bibnamefont {Koma}},\
  }\href@noop {} {\bibfield  {journal} {\bibinfo  {journal} {J. Math. Phys.}\
  }\textbf {\bibinfo {volume} {57}},\ \bibinfo {pages} {021903} (\bibinfo
  {year} {2016})}\BibitemShut {NoStop}%
\bibitem [{\citenamefont {Katsura}\ and\ \citenamefont
  {Koma}(2018)}]{katsura2018noncommutative}%
  \BibitemOpen
  \bibfield  {author} {\bibinfo {author} {\bibfnamefont {H.}~\bibnamefont
  {Katsura}}\ and\ \bibinfo {author} {\bibfnamefont {T.}~\bibnamefont {Koma}},\
  }\href@noop {} {\bibfield  {journal} {\bibinfo  {journal} {J. Math. Phys.}\
  }\textbf {\bibinfo {volume} {59}},\ \bibinfo {pages} {031903} (\bibinfo
  {year} {2018})}\BibitemShut {NoStop}%
\bibitem [{\citenamefont {Odagaki}\ and\ \citenamefont
  {Nguyen}(1986)}]{PhysRevB.33.2184}%
  \BibitemOpen
  \bibfield  {author} {\bibinfo {author} {\bibfnamefont {T.}~\bibnamefont
  {Odagaki}}\ and\ \bibinfo {author} {\bibfnamefont {D.}~\bibnamefont
  {Nguyen}},\ }\href {\doibase 10.1103/PhysRevB.33.2184} {\bibfield  {journal}
  {\bibinfo  {journal} {Phys. Rev. B}\ }\textbf {\bibinfo {volume} {33}},\
  \bibinfo {pages} {2184} (\bibinfo {year} {1986})}\BibitemShut {NoStop}%
\bibitem [{\citenamefont {Liu}\ and\ \citenamefont
  {Ma}(1991)}]{PhysRevB.43.1378}%
  \BibitemOpen
  \bibfield  {author} {\bibinfo {author} {\bibfnamefont {Y.}~\bibnamefont
  {Liu}}\ and\ \bibinfo {author} {\bibfnamefont {P.}~\bibnamefont {Ma}},\
  }\href {\doibase 10.1103/PhysRevB.43.1378} {\bibfield  {journal} {\bibinfo
  {journal} {Phys. Rev. B}\ }\textbf {\bibinfo {volume} {43}},\ \bibinfo
  {pages} {1378} (\bibinfo {year} {1991})}\BibitemShut {NoStop}%
\bibitem [{\citenamefont {Schreiber}(1985)}]{schreiber1985numerical}%
  \BibitemOpen
  \bibfield  {author} {\bibinfo {author} {\bibfnamefont {M.}~\bibnamefont
  {Schreiber}},\ }\href@noop {} {\bibfield  {journal} {\bibinfo  {journal}
  {Journal of Physics C: Solid State Physics}\ }\textbf {\bibinfo {volume}
  {18}},\ \bibinfo {pages} {2493} (\bibinfo {year} {1985})}\BibitemShut
  {NoStop}%
\bibitem [{\citenamefont {Odagaki}(1986)}]{odagaki1986properties}%
  \BibitemOpen
  \bibfield  {author} {\bibinfo {author} {\bibfnamefont {T.}~\bibnamefont
  {Odagaki}},\ }\href@noop {} {\bibfield  {journal} {\bibinfo  {journal} {Solid
  State Commun.}\ }\textbf {\bibinfo {volume} {60}},\ \bibinfo {pages} {693}
  (\bibinfo {year} {1986})}\BibitemShut {NoStop}%
\bibitem [{\citenamefont {de~Laissardi{\`e}re}\ \emph
  {et~al.}(2014)\citenamefont {de~Laissardi{\`e}re}, \citenamefont {Szallas},\
  and\ \citenamefont {Mayou}}]{de2013electronic}%
  \BibitemOpen
  \bibfield  {author} {\bibinfo {author} {\bibfnamefont {G.~T.}\ \bibnamefont
  {de~Laissardi{\`e}re}}, \bibinfo {author} {\bibfnamefont {A.}~\bibnamefont
  {Szallas}}, \ and\ \bibinfo {author} {\bibfnamefont {D.}~\bibnamefont
  {Mayou}},\ }\href@noop {} {\bibfield  {journal} {\bibinfo  {journal} {Acta
  Phys. Pol. A}\ }\textbf {\bibinfo {volume} {126}},\ \bibinfo {pages} {617}
  (\bibinfo {year} {2014})}\BibitemShut {NoStop}%
\bibitem [{\citenamefont {Rieth}\ and\ \citenamefont
  {Schreiber}(1995)}]{PhysRevB.51.15827}%
  \BibitemOpen
  \bibfield  {author} {\bibinfo {author} {\bibfnamefont {T.}~\bibnamefont
  {Rieth}}\ and\ \bibinfo {author} {\bibfnamefont {M.}~\bibnamefont
  {Schreiber}},\ }\href {\doibase 10.1103/PhysRevB.51.15827} {\bibfield
  {journal} {\bibinfo  {journal} {Phys. Rev. B}\ }\textbf {\bibinfo {volume}
  {51}},\ \bibinfo {pages} {15827} (\bibinfo {year} {1995})}\BibitemShut
  {NoStop}%
\bibitem [{\citenamefont {Sutherland}(1986{\natexlab{b}})}]{PhysRevB.34.5208}%
  \BibitemOpen
  \bibfield  {author} {\bibinfo {author} {\bibfnamefont {B.}~\bibnamefont
  {Sutherland}},\ }\href {\doibase 10.1103/PhysRevB.34.5208} {\bibfield
  {journal} {\bibinfo  {journal} {Phys. Rev. B}\ }\textbf {\bibinfo {volume}
  {34}},\ \bibinfo {pages} {5208} (\bibinfo {year}
  {1986}{\natexlab{b}})}\BibitemShut {NoStop}%
\bibitem [{\citenamefont {Choy}(1987)}]{PhysRevB.35.1456}%
  \BibitemOpen
  \bibfield  {author} {\bibinfo {author} {\bibfnamefont {T.~C.}\ \bibnamefont
  {Choy}},\ }\href {\doibase 10.1103/PhysRevB.35.1456} {\bibfield  {journal}
  {\bibinfo  {journal} {Phys. Rev. B}\ }\textbf {\bibinfo {volume} {35}},\
  \bibinfo {pages} {1456} (\bibinfo {year} {1987})}\BibitemShut {NoStop}%
\bibitem [{\citenamefont {Tsunetsugu}\ and\ \citenamefont
  {Ueda}(1988)}]{PhysRevB.38.10109}%
  \BibitemOpen
  \bibfield  {author} {\bibinfo {author} {\bibfnamefont {H.}~\bibnamefont
  {Tsunetsugu}}\ and\ \bibinfo {author} {\bibfnamefont {K.}~\bibnamefont
  {Ueda}},\ }\href {\doibase 10.1103/PhysRevB.38.10109} {\bibfield  {journal}
  {\bibinfo  {journal} {Phys. Rev. B}\ }\textbf {\bibinfo {volume} {38}},\
  \bibinfo {pages} {10109} (\bibinfo {year} {1988})}\BibitemShut {NoStop}%
\bibitem [{\citenamefont {Ledieu}\ \emph {et~al.}(2006)\citenamefont {Ledieu},
  \citenamefont {Unsworth}, \citenamefont {Lograsso}, \citenamefont {Ross},\
  and\ \citenamefont {McGrath}}]{PhysRevB.73.012204}%
  \BibitemOpen
  \bibfield  {author} {\bibinfo {author} {\bibfnamefont {J.}~\bibnamefont
  {Ledieu}}, \bibinfo {author} {\bibfnamefont {P.}~\bibnamefont {Unsworth}},
  \bibinfo {author} {\bibfnamefont {T.~A.}\ \bibnamefont {Lograsso}}, \bibinfo
  {author} {\bibfnamefont {A.~R.}\ \bibnamefont {Ross}}, \ and\ \bibinfo
  {author} {\bibfnamefont {R.}~\bibnamefont {McGrath}},\ }\href {\doibase
  10.1103/PhysRevB.73.012204} {\bibfield  {journal} {\bibinfo  {journal} {Phys.
  Rev. B}\ }\textbf {\bibinfo {volume} {73}},\ \bibinfo {pages} {012204}
  (\bibinfo {year} {2006})}\BibitemShut {NoStop}%
\bibitem [{\citenamefont {Ledieu}\ \emph {et~al.}(2008)\citenamefont {Ledieu},
  \citenamefont {Leung}, \citenamefont {Wearing}, \citenamefont {McGrath},
  \citenamefont {Lograsso}, \citenamefont {Wu},\ and\ \citenamefont
  {Fourn\'ee}}]{PhysRevB.77.073409}%
  \BibitemOpen
  \bibfield  {author} {\bibinfo {author} {\bibfnamefont {J.}~\bibnamefont
  {Ledieu}}, \bibinfo {author} {\bibfnamefont {L.}~\bibnamefont {Leung}},
  \bibinfo {author} {\bibfnamefont {L.~H.}\ \bibnamefont {Wearing}}, \bibinfo
  {author} {\bibfnamefont {R.}~\bibnamefont {McGrath}}, \bibinfo {author}
  {\bibfnamefont {T.~A.}\ \bibnamefont {Lograsso}}, \bibinfo {author}
  {\bibfnamefont {D.}~\bibnamefont {Wu}}, \ and\ \bibinfo {author}
  {\bibfnamefont {V.}~\bibnamefont {Fourn\'ee}},\ }\href {\doibase
  10.1103/PhysRevB.77.073409} {\bibfield  {journal} {\bibinfo  {journal} {Phys.
  Rev. B}\ }\textbf {\bibinfo {volume} {77}},\ \bibinfo {pages} {073409}
  (\bibinfo {year} {2008})}\BibitemShut {NoStop}%
\bibitem [{\citenamefont {Kraj\ifmmode~\check{c}\else \v{c}\fi{}\'{\i}}\ \emph
  {et~al.}(2010)\citenamefont {Kraj\ifmmode~\check{c}\else \v{c}\fi{}\'{\i}},
  \citenamefont {Hafner}, \citenamefont {Ledieu}, \citenamefont {Fourn\'ee},\
  and\ \citenamefont {McGrath}}]{PhysRevB.82.085417}%
  \BibitemOpen
  \bibfield  {author} {\bibinfo {author} {\bibfnamefont {M.}~\bibnamefont
  {Kraj\ifmmode~\check{c}\else \v{c}\fi{}\'{\i}}}, \bibinfo {author}
  {\bibfnamefont {J.}~\bibnamefont {Hafner}}, \bibinfo {author} {\bibfnamefont
  {J.}~\bibnamefont {Ledieu}}, \bibinfo {author} {\bibfnamefont
  {V.}~\bibnamefont {Fourn\'ee}}, \ and\ \bibinfo {author} {\bibfnamefont
  {R.}~\bibnamefont {McGrath}},\ }\href {\doibase 10.1103/PhysRevB.82.085417}
  {\bibfield  {journal} {\bibinfo  {journal} {Phys. Rev. B}\ }\textbf {\bibinfo
  {volume} {82}},\ \bibinfo {pages} {085417} (\bibinfo {year}
  {2010})}\BibitemShut {NoStop}%
\bibitem [{\citenamefont {Ledieu}\ \emph {et~al.}(2009)\citenamefont {Ledieu},
  \citenamefont {Kraj\ifmmode~\check{c}\else \v{c}\fi{}\'{\i}}, \citenamefont
  {Hafner}, \citenamefont {Leung}, \citenamefont {Wearing}, \citenamefont
  {McGrath}, \citenamefont {Lograsso}, \citenamefont {Wu},\ and\ \citenamefont
  {Fourn\'ee}}]{PhysRevB.79.165430}%
  \BibitemOpen
  \bibfield  {author} {\bibinfo {author} {\bibfnamefont {J.}~\bibnamefont
  {Ledieu}}, \bibinfo {author} {\bibfnamefont {M.}~\bibnamefont
  {Kraj\ifmmode~\check{c}\else \v{c}\fi{}\'{\i}}}, \bibinfo {author}
  {\bibfnamefont {J.}~\bibnamefont {Hafner}}, \bibinfo {author} {\bibfnamefont
  {L.}~\bibnamefont {Leung}}, \bibinfo {author} {\bibfnamefont {L.~H.}\
  \bibnamefont {Wearing}}, \bibinfo {author} {\bibfnamefont {R.}~\bibnamefont
  {McGrath}}, \bibinfo {author} {\bibfnamefont {T.~A.}\ \bibnamefont
  {Lograsso}}, \bibinfo {author} {\bibfnamefont {D.}~\bibnamefont {Wu}}, \ and\
  \bibinfo {author} {\bibfnamefont {V.}~\bibnamefont {Fourn\'ee}},\ }\href
  {\doibase 10.1103/PhysRevB.79.165430} {\bibfield  {journal} {\bibinfo
  {journal} {Phys. Rev. B}\ }\textbf {\bibinfo {volume} {79}},\ \bibinfo
  {pages} {165430} (\bibinfo {year} {2009})}\BibitemShut {NoStop}%
\bibitem [{\citenamefont {Deniozou}\ \emph {et~al.}(2009)\citenamefont
  {Deniozou}, \citenamefont {Ledieu}, \citenamefont {Fourn\'ee}, \citenamefont
  {Wu}, \citenamefont {Lograsso}, \citenamefont {Li},\ and\ \citenamefont
  {Diehl}}]{PhysRevB.79.245405}%
  \BibitemOpen
  \bibfield  {author} {\bibinfo {author} {\bibfnamefont {T.}~\bibnamefont
  {Deniozou}}, \bibinfo {author} {\bibfnamefont {J.}~\bibnamefont {Ledieu}},
  \bibinfo {author} {\bibfnamefont {V.}~\bibnamefont {Fourn\'ee}}, \bibinfo
  {author} {\bibfnamefont {D.~M.}\ \bibnamefont {Wu}}, \bibinfo {author}
  {\bibfnamefont {T.~A.}\ \bibnamefont {Lograsso}}, \bibinfo {author}
  {\bibfnamefont {H.~I.}\ \bibnamefont {Li}}, \ and\ \bibinfo {author}
  {\bibfnamefont {R.~D.}\ \bibnamefont {Diehl}},\ }\href {\doibase
  10.1103/PhysRevB.79.245405} {\bibfield  {journal} {\bibinfo  {journal} {Phys.
  Rev. B}\ }\textbf {\bibinfo {volume} {79}},\ \bibinfo {pages} {245405}
  (\bibinfo {year} {2009})}\BibitemShut {NoStop}%
\bibitem [{\citenamefont {Smerdon}\ \emph
  {et~al.}(2008{\natexlab{a}})\citenamefont {Smerdon}, \citenamefont {Leung},
  \citenamefont {Parle}, \citenamefont {Jenks}, \citenamefont {McGrath},
  \citenamefont {Fourn{\'e}e},\ and\ \citenamefont
  {Ledieu}}]{smerdon2008formation}%
  \BibitemOpen
  \bibfield  {author} {\bibinfo {author} {\bibfnamefont {J.}~\bibnamefont
  {Smerdon}}, \bibinfo {author} {\bibfnamefont {L.}~\bibnamefont {Leung}},
  \bibinfo {author} {\bibfnamefont {J.}~\bibnamefont {Parle}}, \bibinfo
  {author} {\bibfnamefont {C.}~\bibnamefont {Jenks}}, \bibinfo {author}
  {\bibfnamefont {R.}~\bibnamefont {McGrath}}, \bibinfo {author} {\bibfnamefont
  {V.}~\bibnamefont {Fourn{\'e}e}}, \ and\ \bibinfo {author} {\bibfnamefont
  {J.}~\bibnamefont {Ledieu}},\ }\href@noop {} {\bibfield  {journal} {\bibinfo
  {journal} {Surf. Sci.}\ }\textbf {\bibinfo {volume} {602}},\ \bibinfo {pages}
  {2496} (\bibinfo {year} {2008}{\natexlab{a}})}\BibitemShut {NoStop}%
\bibitem [{\citenamefont {Sharma}\ \emph {et~al.}(2013)\citenamefont {Sharma},
  \citenamefont {Nozawa}, \citenamefont {Smerdon}, \citenamefont {Nugent},
  \citenamefont {McLeod}, \citenamefont {Dhanak}, \citenamefont {Shimoda},
  \citenamefont {Ishii}, \citenamefont {Tsai},\ and\ \citenamefont
  {McGrath}}]{sharma2013templated}%
  \BibitemOpen
  \bibfield  {author} {\bibinfo {author} {\bibfnamefont {H.}~\bibnamefont
  {Sharma}}, \bibinfo {author} {\bibfnamefont {K.}~\bibnamefont {Nozawa}},
  \bibinfo {author} {\bibfnamefont {J.}~\bibnamefont {Smerdon}}, \bibinfo
  {author} {\bibfnamefont {P.}~\bibnamefont {Nugent}}, \bibinfo {author}
  {\bibfnamefont {I.}~\bibnamefont {McLeod}}, \bibinfo {author} {\bibfnamefont
  {V.}~\bibnamefont {Dhanak}}, \bibinfo {author} {\bibfnamefont
  {M.}~\bibnamefont {Shimoda}}, \bibinfo {author} {\bibfnamefont
  {Y.}~\bibnamefont {Ishii}}, \bibinfo {author} {\bibfnamefont
  {A.}~\bibnamefont {Tsai}}, \ and\ \bibinfo {author} {\bibfnamefont
  {R.}~\bibnamefont {McGrath}},\ }\href@noop {} {\bibfield  {journal} {\bibinfo
   {journal} {Nature Commun.}\ }\textbf {\bibinfo {volume} {4}},\ \bibinfo
  {pages} {2715} (\bibinfo {year} {2013})}\BibitemShut {NoStop}%
\bibitem [{\citenamefont {Sharma}\ \emph {et~al.}(2005)\citenamefont {Sharma},
  \citenamefont {Shimoda}, \citenamefont {Ross}, \citenamefont {Lograsso},\
  and\ \citenamefont {Tsai}}]{PhysRevB.72.045428}%
  \BibitemOpen
  \bibfield  {author} {\bibinfo {author} {\bibfnamefont {H.~R.}\ \bibnamefont
  {Sharma}}, \bibinfo {author} {\bibfnamefont {M.}~\bibnamefont {Shimoda}},
  \bibinfo {author} {\bibfnamefont {A.~R.}\ \bibnamefont {Ross}}, \bibinfo
  {author} {\bibfnamefont {T.~A.}\ \bibnamefont {Lograsso}}, \ and\ \bibinfo
  {author} {\bibfnamefont {A.~P.}\ \bibnamefont {Tsai}},\ }\href {\doibase
  10.1103/PhysRevB.72.045428} {\bibfield  {journal} {\bibinfo  {journal} {Phys.
  Rev. B}\ }\textbf {\bibinfo {volume} {72}},\ \bibinfo {pages} {045428}
  (\bibinfo {year} {2005})}\BibitemShut {NoStop}%
\bibitem [{\citenamefont {Franke}\ \emph {et~al.}(2002)\citenamefont {Franke},
  \citenamefont {Sharma}, \citenamefont {Theis}, \citenamefont {Gille},
  \citenamefont {Ebert},\ and\ \citenamefont {Rieder}}]{PhysRevLett.89.156104}%
  \BibitemOpen
  \bibfield  {author} {\bibinfo {author} {\bibfnamefont {K.~J.}\ \bibnamefont
  {Franke}}, \bibinfo {author} {\bibfnamefont {H.~R.}\ \bibnamefont {Sharma}},
  \bibinfo {author} {\bibfnamefont {W.}~\bibnamefont {Theis}}, \bibinfo
  {author} {\bibfnamefont {P.}~\bibnamefont {Gille}}, \bibinfo {author}
  {\bibfnamefont {P.}~\bibnamefont {Ebert}}, \ and\ \bibinfo {author}
  {\bibfnamefont {K.~H.}\ \bibnamefont {Rieder}},\ }\href {\doibase
  10.1103/PhysRevLett.89.156104} {\bibfield  {journal} {\bibinfo  {journal}
  {Phys. Rev. Lett.}\ }\textbf {\bibinfo {volume} {89}},\ \bibinfo {pages}
  {156104} (\bibinfo {year} {2002})}\BibitemShut {NoStop}%
\bibitem [{\citenamefont {Smerdon}\ \emph
  {et~al.}(2008{\natexlab{b}})\citenamefont {Smerdon}, \citenamefont {Parle},
  \citenamefont {Wearing}, \citenamefont {Lograsso}, \citenamefont {Ross},\
  and\ \citenamefont {McGrath}}]{PhysRevB.78.075407}%
  \BibitemOpen
  \bibfield  {author} {\bibinfo {author} {\bibfnamefont {J.~A.}\ \bibnamefont
  {Smerdon}}, \bibinfo {author} {\bibfnamefont {J.~K.}\ \bibnamefont {Parle}},
  \bibinfo {author} {\bibfnamefont {L.~H.}\ \bibnamefont {Wearing}}, \bibinfo
  {author} {\bibfnamefont {T.~A.}\ \bibnamefont {Lograsso}}, \bibinfo {author}
  {\bibfnamefont {A.~R.}\ \bibnamefont {Ross}}, \ and\ \bibinfo {author}
  {\bibfnamefont {R.}~\bibnamefont {McGrath}},\ }\href {\doibase
  10.1103/PhysRevB.78.075407} {\bibfield  {journal} {\bibinfo  {journal} {Phys.
  Rev. B}\ }\textbf {\bibinfo {volume} {78}},\ \bibinfo {pages} {075407}
  (\bibinfo {year} {2008}{\natexlab{b}})}\BibitemShut {NoStop}%
\bibitem [{\citenamefont {Smerdon}\ \emph
  {et~al.}(2006{\natexlab{a}})\citenamefont {Smerdon}, \citenamefont {Ledieu},
  \citenamefont {Hoeft}, \citenamefont {Reid}, \citenamefont {Wearing},
  \citenamefont {Diehl}, \citenamefont {Lograsso}, \citenamefont {Ross},\ and\
  \citenamefont {McGrath}}]{smerdon2006adsorption}%
  \BibitemOpen
  \bibfield  {author} {\bibinfo {author} {\bibfnamefont {J.~A.}\ \bibnamefont
  {Smerdon}}, \bibinfo {author} {\bibfnamefont {J.}~\bibnamefont {Ledieu}},
  \bibinfo {author} {\bibfnamefont {J.}~\bibnamefont {Hoeft}}, \bibinfo
  {author} {\bibfnamefont {D.}~\bibnamefont {Reid}}, \bibinfo {author}
  {\bibfnamefont {L.~H.}\ \bibnamefont {Wearing}}, \bibinfo {author}
  {\bibfnamefont {R.}~\bibnamefont {Diehl}}, \bibinfo {author} {\bibfnamefont
  {T.}~\bibnamefont {Lograsso}}, \bibinfo {author} {\bibfnamefont
  {A.}~\bibnamefont {Ross}}, \ and\ \bibinfo {author} {\bibfnamefont
  {R.}~\bibnamefont {McGrath}},\ }\href@noop {} {\bibfield  {journal} {\bibinfo
   {journal} {Philos. Mag.}\ }\textbf {\bibinfo {volume} {86}},\ \bibinfo
  {pages} {841} (\bibinfo {year} {2006}{\natexlab{a}})}\BibitemShut {NoStop}%
\bibitem [{\citenamefont {Ledieu}\ \emph {et~al.}(2004)\citenamefont {Ledieu},
  \citenamefont {Hoeft}, \citenamefont {Reid}, \citenamefont {Smerdon},
  \citenamefont {Diehl}, \citenamefont {Lograsso}, \citenamefont {Ross},\ and\
  \citenamefont {McGrath}}]{PhysRevLett.92.135507}%
  \BibitemOpen
  \bibfield  {author} {\bibinfo {author} {\bibfnamefont {J.}~\bibnamefont
  {Ledieu}}, \bibinfo {author} {\bibfnamefont {J.~T.}\ \bibnamefont {Hoeft}},
  \bibinfo {author} {\bibfnamefont {D.~E.}\ \bibnamefont {Reid}}, \bibinfo
  {author} {\bibfnamefont {J.~A.}\ \bibnamefont {Smerdon}}, \bibinfo {author}
  {\bibfnamefont {R.~D.}\ \bibnamefont {Diehl}}, \bibinfo {author}
  {\bibfnamefont {T.~A.}\ \bibnamefont {Lograsso}}, \bibinfo {author}
  {\bibfnamefont {A.~R.}\ \bibnamefont {Ross}}, \ and\ \bibinfo {author}
  {\bibfnamefont {R.}~\bibnamefont {McGrath}},\ }\href {\doibase
  10.1103/PhysRevLett.92.135507} {\bibfield  {journal} {\bibinfo  {journal}
  {Phys. Rev. Lett.}\ }\textbf {\bibinfo {volume} {92}},\ \bibinfo {pages}
  {135507} (\bibinfo {year} {2004})}\BibitemShut {NoStop}%
\bibitem [{\citenamefont {Ledieu}\ \emph {et~al.}(2005)\citenamefont {Ledieu},
  \citenamefont {Hoeft}, \citenamefont {Reid}, \citenamefont {Smerdon},
  \citenamefont {Diehl}, \citenamefont {Ferralis}, \citenamefont {Lograsso},
  \citenamefont {Ross},\ and\ \citenamefont {McGrath}}]{PhysRevB.72.035420}%
  \BibitemOpen
  \bibfield  {author} {\bibinfo {author} {\bibfnamefont {J.}~\bibnamefont
  {Ledieu}}, \bibinfo {author} {\bibfnamefont {J.~T.}\ \bibnamefont {Hoeft}},
  \bibinfo {author} {\bibfnamefont {D.~E.}\ \bibnamefont {Reid}}, \bibinfo
  {author} {\bibfnamefont {J.~A.}\ \bibnamefont {Smerdon}}, \bibinfo {author}
  {\bibfnamefont {R.~D.}\ \bibnamefont {Diehl}}, \bibinfo {author}
  {\bibfnamefont {N.}~\bibnamefont {Ferralis}}, \bibinfo {author}
  {\bibfnamefont {T.~A.}\ \bibnamefont {Lograsso}}, \bibinfo {author}
  {\bibfnamefont {A.~R.}\ \bibnamefont {Ross}}, \ and\ \bibinfo {author}
  {\bibfnamefont {R.}~\bibnamefont {McGrath}},\ }\href {\doibase
  10.1103/PhysRevB.72.035420} {\bibfield  {journal} {\bibinfo  {journal} {Phys.
  Rev. B}\ }\textbf {\bibinfo {volume} {72}},\ \bibinfo {pages} {035420}
  (\bibinfo {year} {2005})}\BibitemShut {NoStop}%
\bibitem [{\citenamefont {Smerdon}\ \emph
  {et~al.}(2006{\natexlab{b}})\citenamefont {Smerdon}, \citenamefont {Ledieu},
  \citenamefont {McGrath}, \citenamefont {Noakes}, \citenamefont {Bailey},
  \citenamefont {Draxler}, \citenamefont {McConville}, \citenamefont
  {Lograsso},\ and\ \citenamefont {Ross}}]{PhysRevB.74.035429}%
  \BibitemOpen
  \bibfield  {author} {\bibinfo {author} {\bibfnamefont {J.~A.}\ \bibnamefont
  {Smerdon}}, \bibinfo {author} {\bibfnamefont {J.}~\bibnamefont {Ledieu}},
  \bibinfo {author} {\bibfnamefont {R.}~\bibnamefont {McGrath}}, \bibinfo
  {author} {\bibfnamefont {T.~C.~Q.}\ \bibnamefont {Noakes}}, \bibinfo {author}
  {\bibfnamefont {P.}~\bibnamefont {Bailey}}, \bibinfo {author} {\bibfnamefont
  {M.}~\bibnamefont {Draxler}}, \bibinfo {author} {\bibfnamefont {C.~F.}\
  \bibnamefont {McConville}}, \bibinfo {author} {\bibfnamefont {T.~A.}\
  \bibnamefont {Lograsso}}, \ and\ \bibinfo {author} {\bibfnamefont {A.~R.}\
  \bibnamefont {Ross}},\ }\href {\doibase 10.1103/PhysRevB.74.035429}
  {\bibfield  {journal} {\bibinfo  {journal} {Phys. Rev. B}\ }\textbf {\bibinfo
  {volume} {74}},\ \bibinfo {pages} {035429} (\bibinfo {year}
  {2006}{\natexlab{b}})}\BibitemShut {NoStop}%
\bibitem [{\citenamefont {F{\"o}rster}\ \emph {et~al.}(2013)\citenamefont
  {F{\"o}rster}, \citenamefont {Meinel}, \citenamefont {Hammer}, \citenamefont
  {Trautmann},\ and\ \citenamefont {Widdra}}]{forster2013quasicrystalline}%
  \BibitemOpen
  \bibfield  {author} {\bibinfo {author} {\bibfnamefont {S.}~\bibnamefont
  {F{\"o}rster}}, \bibinfo {author} {\bibfnamefont {K.}~\bibnamefont {Meinel}},
  \bibinfo {author} {\bibfnamefont {R.}~\bibnamefont {Hammer}}, \bibinfo
  {author} {\bibfnamefont {M.}~\bibnamefont {Trautmann}}, \ and\ \bibinfo
  {author} {\bibfnamefont {W.}~\bibnamefont {Widdra}},\ }\href@noop {}
  {\bibfield  {journal} {\bibinfo  {journal} {Nature}\ }\textbf {\bibinfo
  {volume} {502}},\ \bibinfo {pages} {215} (\bibinfo {year}
  {2013})}\BibitemShut {NoStop}%
\bibitem [{\citenamefont {Smith}\ \emph {et~al.}(1996)\citenamefont {Smith},
  \citenamefont {Chao}, \citenamefont {Niu},\ and\ \citenamefont
  {Shih}}]{ScienceAgGaAs}%
  \BibitemOpen
  \bibfield  {author} {\bibinfo {author} {\bibfnamefont {A.~R.}\ \bibnamefont
  {Smith}}, \bibinfo {author} {\bibfnamefont {K.-J.}\ \bibnamefont {Chao}},
  \bibinfo {author} {\bibfnamefont {Q.}~\bibnamefont {Niu}}, \ and\ \bibinfo
  {author} {\bibfnamefont {C.-K.}\ \bibnamefont {Shih}},\ }\href@noop {}
  {\bibfield  {journal} {\bibinfo  {journal} {Science}\ }\textbf {\bibinfo
  {volume} {273}},\ \bibinfo {pages} {226} (\bibinfo {year}
  {1996})}\BibitemShut {NoStop}%
\bibitem [{\citenamefont {Collins}\ \emph {et~al.}(2017)\citenamefont
  {Collins}, \citenamefont {Witte}, \citenamefont {Silverman}, \citenamefont
  {Green},\ and\ \citenamefont {Gomes}}]{collins2017imaging}%
  \BibitemOpen
  \bibfield  {author} {\bibinfo {author} {\bibfnamefont {L.~C.}\ \bibnamefont
  {Collins}}, \bibinfo {author} {\bibfnamefont {T.~G.}\ \bibnamefont {Witte}},
  \bibinfo {author} {\bibfnamefont {R.}~\bibnamefont {Silverman}}, \bibinfo
  {author} {\bibfnamefont {D.~B.}\ \bibnamefont {Green}}, \ and\ \bibinfo
  {author} {\bibfnamefont {K.~K.}\ \bibnamefont {Gomes}},\ }\href@noop {}
  {\bibfield  {journal} {\bibinfo  {journal} {Nature Commun.}\ }\textbf
  {\bibinfo {volume} {8}},\ \bibinfo {pages} {15961} (\bibinfo {year}
  {2017})}\BibitemShut {NoStop}%
\bibitem [{\citenamefont {Senechal}(1996)}]{senechal1996quasicrystals}%
  \BibitemOpen
  \bibfield  {author} {\bibinfo {author} {\bibfnamefont {M.}~\bibnamefont
  {Senechal}},\ }\href@noop {} {\emph {\bibinfo {title} {Quasicrystals and
  geometry}}}\ (\bibinfo  {publisher} {Cambridge University Press},\ \bibinfo
  {year} {1996})\BibitemShut {NoStop}%
\bibitem [{\citenamefont {Zhang}\ and\ \citenamefont
  {Zhang}(2013)}]{zhang2013topological}%
  \BibitemOpen
  \bibfield  {author} {\bibinfo {author} {\bibfnamefont {H.}~\bibnamefont
  {Zhang}}\ and\ \bibinfo {author} {\bibfnamefont {S.-C.}\ \bibnamefont
  {Zhang}},\ }\href@noop {} {\bibfield  {journal} {\bibinfo  {journal} {Phys.
  Status Solidi (RRL)}\ }\textbf {\bibinfo {volume} {7}},\ \bibinfo {pages}
  {72} (\bibinfo {year} {2013})}\BibitemShut {NoStop}%
\bibitem [{\citenamefont {Liang}\ \emph {et~al.}(2016)\citenamefont {Liang},
  \citenamefont {Yu}, \citenamefont {Zhou},\ and\ \citenamefont
  {Hu}}]{PhysRevB.93.035135}%
  \BibitemOpen
  \bibfield  {author} {\bibinfo {author} {\bibfnamefont {Q.-F.}\ \bibnamefont
  {Liang}}, \bibinfo {author} {\bibfnamefont {R.}~\bibnamefont {Yu}}, \bibinfo
  {author} {\bibfnamefont {J.}~\bibnamefont {Zhou}}, \ and\ \bibinfo {author}
  {\bibfnamefont {X.}~\bibnamefont {Hu}},\ }\href {\doibase
  10.1103/PhysRevB.93.035135} {\bibfield  {journal} {\bibinfo  {journal} {Phys.
  Rev. B}\ }\textbf {\bibinfo {volume} {93}},\ \bibinfo {pages} {035135}
  (\bibinfo {year} {2016})}\BibitemShut {NoStop}%
\bibitem [{\citenamefont {Zhang}\ \emph {et~al.}(2011)\citenamefont {Zhang},
  \citenamefont {Yu}, \citenamefont {Feng}, \citenamefont {Yao}, \citenamefont
  {Weng}, \citenamefont {Dai},\ and\ \citenamefont
  {Fang}}]{PhysRevLett.106.156808}%
  \BibitemOpen
  \bibfield  {author} {\bibinfo {author} {\bibfnamefont {W.}~\bibnamefont
  {Zhang}}, \bibinfo {author} {\bibfnamefont {R.}~\bibnamefont {Yu}}, \bibinfo
  {author} {\bibfnamefont {W.}~\bibnamefont {Feng}}, \bibinfo {author}
  {\bibfnamefont {Y.}~\bibnamefont {Yao}}, \bibinfo {author} {\bibfnamefont
  {H.}~\bibnamefont {Weng}}, \bibinfo {author} {\bibfnamefont {X.}~\bibnamefont
  {Dai}}, \ and\ \bibinfo {author} {\bibfnamefont {Z.}~\bibnamefont {Fang}},\
  }\href {\doibase 10.1103/PhysRevLett.106.156808} {\bibfield  {journal}
  {\bibinfo  {journal} {Phys. Rev. Lett.}\ }\textbf {\bibinfo {volume} {106}},\
  \bibinfo {pages} {156808} (\bibinfo {year} {2011})}\BibitemShut {NoStop}%
\bibitem [{\citenamefont {Si}\ \emph {et~al.}(2016)\citenamefont {Si},
  \citenamefont {Jin}, \citenamefont {Zhou}, \citenamefont {Sun},\ and\
  \citenamefont {Liu}}]{si2016large}%
  \BibitemOpen
  \bibfield  {author} {\bibinfo {author} {\bibfnamefont {C.}~\bibnamefont
  {Si}}, \bibinfo {author} {\bibfnamefont {K.-H.}\ \bibnamefont {Jin}},
  \bibinfo {author} {\bibfnamefont {J.}~\bibnamefont {Zhou}}, \bibinfo {author}
  {\bibfnamefont {Z.}~\bibnamefont {Sun}}, \ and\ \bibinfo {author}
  {\bibfnamefont {F.}~\bibnamefont {Liu}},\ }\href@noop {} {\bibfield
  {journal} {\bibinfo  {journal} {Nano Lett.}\ }\textbf {\bibinfo {volume}
  {16}},\ \bibinfo {pages} {6584} (\bibinfo {year} {2016})}\BibitemShut
  {NoStop}%
\bibitem [{\citenamefont {Dzero}\ \emph {et~al.}(2010)\citenamefont {Dzero},
  \citenamefont {Sun}, \citenamefont {Galitski},\ and\ \citenamefont
  {Coleman}}]{PhysRevLett.104.106408}%
  \BibitemOpen
  \bibfield  {author} {\bibinfo {author} {\bibfnamefont {M.}~\bibnamefont
  {Dzero}}, \bibinfo {author} {\bibfnamefont {K.}~\bibnamefont {Sun}}, \bibinfo
  {author} {\bibfnamefont {V.}~\bibnamefont {Galitski}}, \ and\ \bibinfo
  {author} {\bibfnamefont {P.}~\bibnamefont {Coleman}},\ }\href {\doibase
  10.1103/PhysRevLett.104.106408} {\bibfield  {journal} {\bibinfo  {journal}
  {Phys. Rev. Lett.}\ }\textbf {\bibinfo {volume} {104}},\ \bibinfo {pages}
  {106408} (\bibinfo {year} {2010})}\BibitemShut {NoStop}%
\bibitem [{\citenamefont {Levine}\ and\ \citenamefont
  {Steinhardt}(1986)}]{PhysRevB.34.596}%
  \BibitemOpen
  \bibfield  {author} {\bibinfo {author} {\bibfnamefont {D.}~\bibnamefont
  {Levine}}\ and\ \bibinfo {author} {\bibfnamefont {P.~J.}\ \bibnamefont
  {Steinhardt}},\ }\href {\doibase 10.1103/PhysRevB.34.596} {\bibfield
  {journal} {\bibinfo  {journal} {Phys. Rev. B}\ }\textbf {\bibinfo {volume}
  {34}},\ \bibinfo {pages} {596} (\bibinfo {year} {1986})}\BibitemShut
  {NoStop}%
\bibitem [{\citenamefont {Socolar}\ and\ \citenamefont
  {Steinhardt}(1986)}]{PhysRevB.34.617}%
  \BibitemOpen
  \bibfield  {author} {\bibinfo {author} {\bibfnamefont {J.~E.~S.}\
  \bibnamefont {Socolar}}\ and\ \bibinfo {author} {\bibfnamefont {P.~J.}\
  \bibnamefont {Steinhardt}},\ }\href {\doibase 10.1103/PhysRevB.34.617}
  {\bibfield  {journal} {\bibinfo  {journal} {Phys. Rev. B}\ }\textbf {\bibinfo
  {volume} {34}},\ \bibinfo {pages} {617} (\bibinfo {year} {1986})}\BibitemShut
  {NoStop}%
\end{thebibliography}
\providecommand{\noopsort}[1]{}\providecommand{\singleletter}[1]{#1}%

\end{document}